\documentclass[amsmath,amssymb,floatfix,aps,prl,reprint,
               twocolumn,superscriptaddress,nobalancelastpage
               ]{revtex4-1}

\usepackage{graphicx}
\usepackage{dcolumn}
\usepackage{bm}
\usepackage{amsmath}
\usepackage[colorlinks,allcolors=blue]{hyperref}

\graphicspath{ {./} }

\begin{document}

\title{Next-to-Leading Order \textit{Ab Initio} Electron-Phonon Scattering}

\author{Nien-En Lee}
\affiliation{Department of Applied Physics and Materials Science, California Institute of Technology, Pasadena, California 91125, USA}
\affiliation{Department of Physics, California Institute of Technology, Pasadena, California 91125, USA}

\author{Jin-Jian Zhou}
\affiliation{Department of Applied Physics and Materials Science, California Institute of Technology, Pasadena, California 91125, USA}

\author{Hsiao-Yi Chen}
\affiliation{Department of Applied Physics and Materials Science, California Institute of Technology, Pasadena, California 91125, USA}
\affiliation{Department of Physics, California Institute of Technology, Pasadena, California 91125, USA}

\author{Marco Bernardi}
\affiliation{Department of Applied Physics and Materials Science, California Institute of Technology, Pasadena, California 91125, USA}

\date{\today}

\begin{abstract} 
\noindent
Electron-phonon ($e$-ph) interactions are usually treated in the lowest order of perturbation theory.   
Here we derive next-to-leading order $e$-ph interactions, and compute from first principles the associated two-phonon $e$-ph scattering rates. 
The derivation involves Matsubara sums of the relevant two-loop Feynman diagrams, 
and the numerical calculations are challenging since they involve Brillouin zone integrals over two crystal momenta and depend critically on the intermediate state lifetimes. 
Using random grids and Monte Carlo integration, together with a self-consistent update of the intermediate state lifetimes, we compute and converge the two-phonon scattering rates, using GaAs as a case study. 
For the longitudinal optical phonon in GaAs, we find that the two-phonon scattering rates are as large as nearly half the value of the leading-order rates. 
The energy and temperature dependence of the two-phonon processes are analyzed. 
We show that including the two-phonon processes is important to accurately predicting the electron mobility in GaAs. 
\end{abstract} 


\maketitle
\begin{figure}[b]
\includegraphics[scale=0.85]{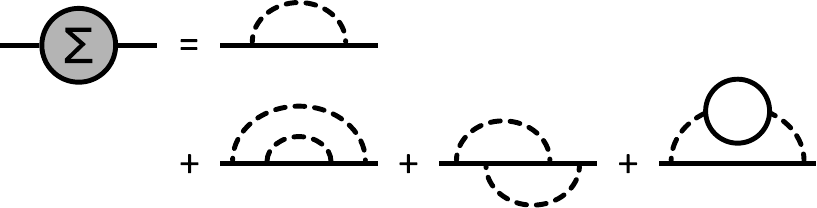}
\caption{Diagrams for the $e$-ph self-energy up to $\mathcal{O}$($g^4$), where $g$ is the $e$-ph coupling constant. 
The first two diagrams in the second row contribute to the two-phonon scattering processes.} 
\label{fig:electron_self_energy}
\end{figure}
\begin{figure}[!t]
\includegraphics[scale=0.9]{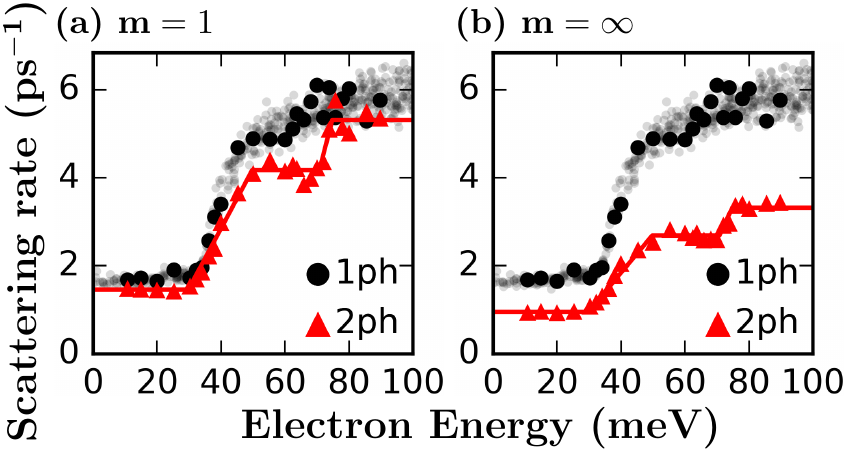}
\caption{Calculated 2ph scattering rates, $\Gamma^{(\textrm{2ph})}_{n\mathbf{k}}$ in Eq.~(\ref{eq:1}), for electrons in GaAs at 300 K. The zero of the energy axis is the conduction band minimum. 
(a) Shows the first iteration and (b) the final result after converging the intermediate lifetime update procedure. 
The lowest-order $e$-ph scattering rates, denoted as 1ph, are also given for comparison.}
%The red line segments are the fitting of $\Gamma^{(\textrm{2ph})}$ used in the computation for the next iteration. Black circles are their corresponding first order $e$-ph scattering rates, and the background black dots are that on a fine \textbf{k}-point grid.} 
\label{fig:total_2ph_rate}
\end{figure}

Electron-phonon ($e$-ph) interactions are essential to understanding electrical transport, nonequilibrium dynamics and superconductivity. 
Using density functional theory (DFT) and related methods, it has become possible to compute $e$-ph interactions from first principles, 
and use them to predict the carrier scattering rates and mobilities,  
both in simple and in complex materials with up to tens of atoms in the unit cell
~\cite{Bernardi-review, Louie2014, Bernardi-noble, BernardiGaAs, Zhou2017, Chen2017, BernardiSTO, Bernardi-naph, Li2015, Li2018, Sohier2018}. 
In the typical workflow, one takes into account only the leading order $e$-ph scattering processes, which involve scattering of the carriers with one phonon. 
Nearly all work to date has relied on such leading-order perturbation theory, tacitly neglecting higher-order $e$-ph processes.\\
\indent 
Yet, it is well known that many compounds $-$ including polar materials, oxides and organic crystals $-$ exhibit intermediate to strong $e$-ph interactions, which cannot be described within lowest-order theory. 
In the weak to intermediate $e$-ph coupling regime, one expects that perturbation theory is still valid, but that higher-order processes are significant and need to be included; 
in the strong coupling limit, the $e$-ph interactions can lead to regimes beyond the reach of perturbation theory, including the formation and trapping of polarons. Studies of second and even third order $e$-ph corrections exist~\cite{Mahan2000,Smondyrev1986}, but they are limited to simplified models, 
restricted to the conduction band minimum, or only valid at zero temperature, and are therefore inadequate for accurate predictions. 
The intermediate $e$-ph coupling regime has also been investigated using diagram resummation techniques such as the cumulant method, both analytically~\cite{Schonhammer1994} and more recently \textit{ab initio}~\cite{Rehr2014, Gonze2018}.\\  
\indent
Higher-order processes are important in quantum field theories (QFT) of condensed matter. An example are light-matter interactions, where phonon-assisted~\cite{Cohen2012} and two-photon~\cite{Murayama1995, Hayat2008} absorption have been studied extensively. Going beyond the leading order in QFT can provide important corrections $-$ in electron-positron collisions, for example, 
higher order corrections are essential to accurately predicting large-angle Bhabha scattering~\cite{Bern}.
However, it is a daunting task to systematically go beyond the leading order due to the rapid increase in the number of Feynman diagrams and their numerical complexity. 
For the $e$-ph interactions, the next-to-leading order diagrams (see Fig.~\ref{fig:electron_self_energy}) have not been studied in detail $-$ 
their analytic expression has not yet been derived, let alone numerically computed from first principles.\\
%
% Figure 1
%
%
% Figure 2
% 
\begin{figure*}[!t]
\includegraphics[scale=0.95]{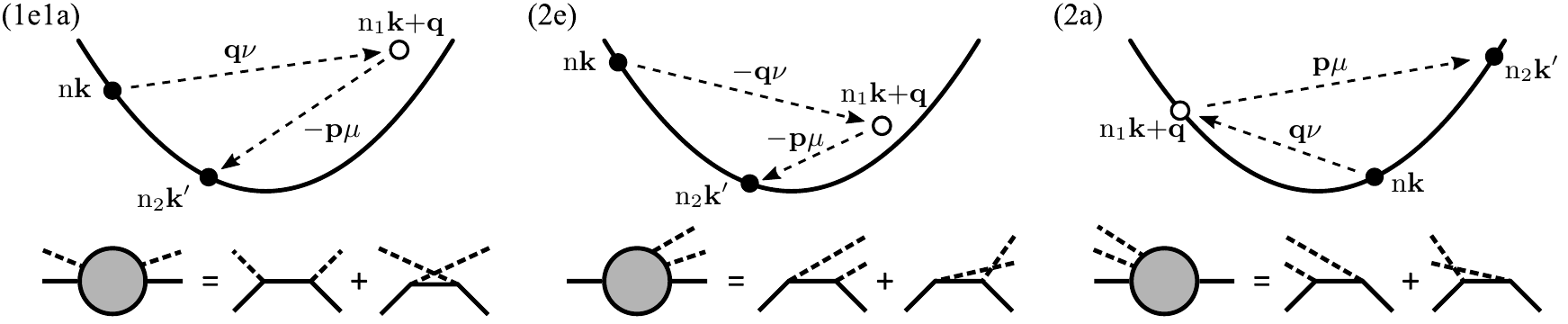}
\caption{$e$-ph scattering processes with two external phonon lines. Shown are the three relevant processes $-$ one phonon absorption plus one phonon emission (left panel, labelled 1e1a), two-phonon emission (middle panel, labelled 2e) and two-phonon absorption (right panel, labelled 2a). Each of these three processes comprises two interfering scattering channels, only one of which is shown in the band structure schematics. Note that the intermediate state does not need to be on-shell.}%, that is, its energy does not necessarily equal a band energy.}%, as shown in the left two panels. } 
\label{fig:two_phonon_processes}
\end{figure*}
\indent
Here we formulate and compute from first principles next-to-leading order $e$-ph interactions, focusing on $e$-ph processes involving two phonons (hereafter denoted as 2ph processes; see Fig. \ref{fig:two_phonon_processes}). We compute and analyze their contributions to the $e$-ph scattering rates, using GaAs as a case study. 
Surprisingly, we find that the 2ph scattering rates are comparable in magnitude to the lowest order rates due to one-phonon processes. 
Our analysis shows that the relative importance of the 2ph contributions is nearly temperature independent in the 200$-$400 K range, 
and rationalizes the peculiar dependence on electron energy of the 2ph processes. 
The results are sensitive to the lifetimes of the intermediate states, which need to be included to avoid divergences due to resonance effects; 
we develop a self-consistent scheme to overcome this challenge. 
The 2ph processes are also shown to play an important role in accurately predicting the electron mobility in GaAs. We formulate and iteratively solve a linearized 
Boltzmann transport equation (BTE) that correctly includes the 2ph processes, showing that this level of theory can correct the discrepancy with experiment 
of the mobility predicted with the BTE including only one-phonon processes. 
Our work provides a framework for systematically improving the accuracy of \textit{ab initio} $e$-ph calculations beyond the leading order.\\
\indent
We use the Matsubara frequency sum technique to derive an analytic expression for the electron self-energy due to 
$e$-ph interactions up to $\mathcal{O}$($g^4$) (see Fig. \ref{fig:electron_self_energy}), where $g$ is the $e$-ph coupling constant; our treatment   
focuses on the imaginary part of the self-energy and the related 2ph scattering rates. The derivations are lengthy and tedious, and are given in detail in the Supplemental Material~\cite{SupplementalMaterial}. 
The key result is the scattering rate due to the 2ph processes, $\Gamma^{\textrm{(2ph)}}_{n\textbf{k}}$, for an electronic state with band index $n$ and crystal momentum $\mathbf{k}$, 
which can be expressed as
\begin{align}
\Gamma^{\textrm{(2ph)}}_{n\textbf{k}} =
       \frac{2\pi}{\hbar} \frac{1}{N_{\Omega}^{2}}\;
       \sum_{n_2}\;
       \sum_{\nu\textbf{q}}\;
       \sum_{\mu\textbf{p}}
       \left[\;
           \widetilde{\Gamma}^{\textrm{(1e1a)}}
          +\widetilde{\Gamma}^{\textrm{(2e)}}
          +\widetilde{\Gamma}^{\textrm{(2a)}}
       \right],
       %\nonumber
\label{eq:1}      
\end{align}
where the process-resolved 2ph scattering rates $\widetilde{\Gamma}^{(\rm{i})}$, which depend on the two phonon momenta $\textbf{q}$ and $\textbf{p}$ and their respective mode indexes $\nu$ and $\mu$, 
are labelled (1e1a) for processes that absorb one phonon and emit another phonon, or vice versa, (2e) for processes that emit two phonons, and (2a) for processes that absorb two phonons (see Fig. \ref{fig:two_phonon_processes}). 
In addition, $n_2$ is the band index of the final electronic state, whose momentum is fixed to $\textbf{k}^{\prime}\equiv\textbf{k}+\textbf{q}+\textbf{p}$ by momentum conservation, 
and to correctly normalize the sum, we divide it by $N_{\Omega}^2$, which is the number of points sampled in the ($\textbf{q}$, $\textbf{p}$) space. 
The process-resolved 2ph scattering rates are defined as 
\begin{align}
\label{eq:Gamma_abc}
\widetilde{\Gamma}^{\textrm{(i)}} = 
           \gamma^{\textrm{(i)}}\;
           \delta (
               \varepsilon_{n\textbf{k}}
               -\varepsilon_{n_2\textbf{k}^{\prime}}
               -\alpha_{\bf{p}}^{\textrm{(i)}} 
                   \omega_{\mu\textbf{p}}
               -\alpha_{\bf{q}}^{\textrm{(i)}} 
                   \omega_{\nu\textbf{q}}
           ),
\end{align}
where $\varepsilon$ denotes electron energies relative to the chemical potential, and $\omega$ phonon energies;  
the delta function imposes energy conservation, and the constants $\alpha$ for each process are defined as
%
%
% NUMERICAL FACTORS
\begin{align}
\alpha_{\bf{p}}^{\textrm{(1e1a)}}= \;\;\: 1&,\;\;
\alpha_{\bf{p}}^{\textrm{(2e)}}=  1,\;\;
\alpha_{\bf{p}}^{\textrm{(2a)}}= -1, \nonumber\\
\alpha_{\bf{q}}^{\textrm{(1e1a)}}= -1&,\;\;
\alpha_{\bf{q}}^{\textrm{(2e)}}=  1,\;\;
\alpha_{\bf{q}}^{\textrm{(2a)}}= -1. \nonumber
\end{align}
%
%
% AMPLITUDES
The square amplitudes of the three processes are 
\begin{align}
\label{eq:lowercase_gamma_abc}
\gamma^{\textrm{(i)}} = &\; A^{\textrm{(i)}} \times \\
   \;&\left|
   \sum_{n_1}
   \left(
       \frac{g_{n_1n\nu}{(\bf{k},\bf{q})}
             g_{n_2n_1\mu}(\bf{k}+\bf{q},\bf{p})}
       {\varepsilon_{n_2\textbf{k}^{\prime}}
        -\varepsilon_{n_{1}\textbf{k}+\textbf{q}}
        +\alpha_{\bf{p}}^{\textrm{(i)}}\omega_{\mu\bf{p}}
        +i\eta
        -\Sigma_{n_{1}\textbf{k}+\textbf{q}}
        }
   \right.\right.\nonumber\\
  &+
   \left.\left.
       \frac{g_{n_1n\mu}{(\bf{k},\bf{p})}
             g_{n_2n_1\nu}(\bf{k}+\bf{p},\bf{q})}
       {\varepsilon_{n_2\textbf{k}^{\prime}}
        -\varepsilon_{n_{1}\textbf{k}+\textbf{p}}
        +\alpha_{\bf{q}}^{\textrm{(i)}}\omega_{\nu\bf{q}}
        +i\eta
        -\Sigma_{n_{1}\textbf{k}+\textbf{p}}
        }
   \right)\right|^2
   \nonumber,
\end{align}
where $n_1$ is the band index and $\Sigma$ the self-energy of the intermediate electronic state, and $\eta$ is a positive infinitesimal. 
% PREFACTORS
The prefactors $A^{\textrm{(i)}}$ contain the thermal occupation numbers of electrons and phonons (denoted by $f$ and $N$, respectively) and are defined as:
\begin{align}
\label{eq:thermal_factor_A}
A^{\textrm{(1e1a)}} = &\; 
        N_{\nu\bf{q}}+N_{\nu\bf{q}}N_{\mu\bf{p}}
        +N_{\mu\bf{p}}f_{n_2\textbf{k}^{\prime}}
        -N_{\nu\bf{q}}f_{n_2\textbf{k}^{\prime}}
    \nonumber,
    \\[2ex]
A^{\textrm{(2e)}} = &\; 
    \frac{1}{2}\left[
        (1+N_{\nu\bf{q}})
        (1+N_{\mu\bf{p}}-f_{n_2\textbf{k}^{\prime}})
        -N_{\mu\bf{p}}f_{n_2\textbf{k}^{\prime}} 
    \right]
    \nonumber,
    \\[2ex]
A^{\textrm{(2a)}} = &\; 
    \frac{1}{2}\left[
        N_{\nu\bf{q}}
        (N_{\mu\bf{p}}+f_{n_2\textbf{k}^{\prime}})
        +(1+N_{\mu\bf{p}})f_{n_2\textbf{k}^{\prime}}
    \right]
    .
\end{align}
%
% INTERMEDIATE STATE
%
Note that the intermediate electronic state in the 2ph processes can be off-shell, namely its energy does not need to correspond to an electronic eigenstate, as is shown schematically 
for the (1e1a) and (2e) processes in Fig. \ref{fig:two_phonon_processes}. 
When the intermediate state is on shell, as in the (2a) process in Fig. \ref{fig:two_phonon_processes}, the scattering process results in resonance effects, 
and the intermediate state lifetimes are crucial to prevent the 2ph scattering rates from diverging. 
Here and below, the intermediate state lifetime is defined as the inverse of the scattering rate of the intermediate state, which is obtained from the imaginary part of the intermediate state self-energy as $2/\hbar\, \mathrm{Im}\Sigma$.\\
% SELF-CONSISTENT
%
% NUMERICAL CALCULATIONS
\indent 
The numerical calculations on GaAs follow our previous work~\citep{BernardiGaAs}. Briefly, we carry out DFT calculations on GaAs using the {\sc Quantum ESPRESSO} code \cite{QE-2009} with a plane-wave basis set. We employ the local density approximation \cite{LDA1981} and norm-conserving pseudopotentials \cite{NormCon1991}. 
A relaxed lattice constant of 5.55 \AA, a kinetic energy cutoff of 72 Ry and $8 \times 8 \times 8$ \textbf{k}-point grids are used in all DFT calculations.
% DFPT, $e$-ph
Phonon dispersions are computed with density functional perturbation theory (DFPT) \cite{Giannozzi2001} on an $8 \times 8 \times 8$ \textbf{q}-point grid. 
The $e$-ph coupling constants, $g_{nm\nu}(\textbf{k},\textbf{q})$, are computed on coarse \textbf{k}- and \textbf{q}-point grids~\cite{Bernardi-review} 
using DFPT together with our in-house developed \textsc{Perturbo} code \cite{[The code employed in this work will be released in the future at ]PerturboWebsite} and interpolated using Wannier functions \cite{Giustino2007} generated with the Wannier90 code~\cite{Marzari2014}. 
% LO ONLY
Since in GaAs the longitudinal optical (LO) phonon dominates $e$-ph scattering for electrons within $\sim$100 meV of the conduction band edge~\cite{BernardiGaAs}, 
in the 2ph scattering rate calculations we use only the long-range coupling to the LO modes~\cite{Mauri2015,Giustino2015,Vogl1976} and neglect the coupling to all other modes; 
we have checked that this approximation does not affect the results.
To compute and converge the 2ph scattering rates, we use Monte Carlo integration by sampling up to 3 billion random (\textbf{q},\,\textbf{p}) pairs of Brillouin zone points 
drawn from a Cauchy distribution~\cite{BernardiGaAs}. 
The delta function in Eq. (\ref{eq:Gamma_abc}) is approximated by a Gaussian with a small broadening of $5$ meV.\\ 
\indent 
The 2ph scattering rates are sensitive to the value of the intermediate state self-energy, $\Sigma$ in the denominator of Eq. (\ref{eq:lowercase_gamma_abc}), 
whose value needs to be chosen carefully. We neglect the real part of the intermediate state self-energy, which only corrects the band structure and barely affects the 2ph calculation. 
The imaginary part of $\Sigma$ includes in principle scattering effects from all possible sources. 
In practice, we approximate $\rm{Im}\Sigma$ with the total $e$-ph scattering rate, including both the lowest order and the 2ph rates, using $\left| \mathrm{Im}\Sigma \right| = \hbar/2 [\Gamma^{(\textrm{1ph})} + \Gamma^{(\textrm{2ph})}]$, where $\Gamma^{(\textrm{1ph})}$ is the usual leading-order $e$-ph scattering rate~\cite{BernardiGaAs}. 
This approach makes Eq.~(\ref{eq:1}) a self-consistent problem. 
We iterate Eq.~(\ref{eq:1}) until the 2ph scattering rates equal the 2ph contribution to the intermediate state lifetime.  
In each iteration $m$, the lifetime is due to the sum of the lowest order plus the 2ph scattering rates at the previous iteration, 
namely, $\left| \mathrm{Im}\Sigma (m) \right| = \hbar/2 \, [ \Gamma^{(\textrm{1ph})} + \Gamma^{(\textrm{2ph})}(m-1)]$. 
The initial $\Gamma^{(\textrm{2ph})}$ is set to zero, and the convergence process is performed separately at each temperature.\\ 
\indent 
Figure~\ref{fig:total_2ph_rate} shows the first iteration and the converged result for the 2ph scattering rates in GaAs at 300 K, for states near the bottom of the conduction band. 
The converged 2ph scattering rates are surprisingly large $-$ they are smaller than the leading-order 1ph rate, thus justifying the perturbative approach, but they are nearly half the value of the 1ph rates at all energies. It is clear therefore that LO phonons couple strongly to electrons in GaAs, and that lowest-order perturbation theory can capture only part of the dynamical $e$-ph processes.\\ 
\indent
%
% FIGURE 3
%
%
% FIGURE 4
%
\begin{figure}[!b]
\includegraphics[scale=0.80]{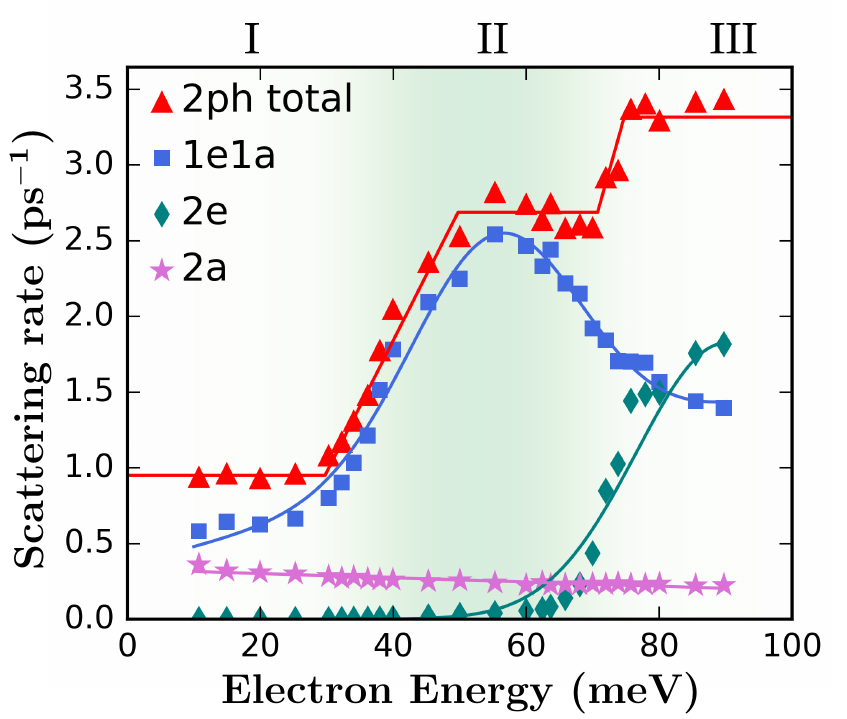}
\caption{Contributions to the 2ph scattering rate in GaAs at 300 K. The scattering rates of the 1e1a, 2e and 2a processes are shown, together with their sum, the total 2ph scattering rate. 
The solid curves fit the data and aid their visualization.}
\label{fig:process_resolved_2ph}
\end{figure}
The energy scale of the LO mode, $\omega_{\mathrm{LO}}\!\approx \!35$~meV, defines three energy regions near the bottom of the conduction band. 
We reference hereafter all electron energies to the conduction band minimum, which is taken to be the energy zero.  
The three energy regions, denoted as I, II and III in Fig.~\ref{fig:process_resolved_2ph}, correspond 
to electron energies below $\omega_{\textrm{LO}}$, between $\omega_{\textrm{LO}}$ and $2\omega_{\textrm{LO}}$, and greater than $2\omega_{\textrm{LO}}$, respectively. 
The 2ph scattering rates exhibit a trend as a function of energy with \textit{three plateaus}, one in each energy region, as seen in Fig.~\ref{fig:process_resolved_2ph}. %determined by the LO mode energy. 
This trend can be understood by plotting the 1e1a, 2e and 2a contributions (see Fig.~\ref{fig:process_resolved_2ph}) 
and analyzing the scattering phase space for emitting or absorbing LO phonons in each energy region.\\
\indent 
In region I, electrons possess an energy smaller than $\omega_{\textrm{LO}}$, and thus cannot emit one or two LO phonons, since this would require a final state in the band gap. 
The 1e1a process is also suppressed in region I since its intermediate state, resulting from one phonon emission, is always virtual. 
The two-phonon absorption (2a) process is active in region I, but it is thermally activated and thus weak at 300 K since $\omega_{\mathrm{LO}}\!\approx \!35$~meV. 
The (2a) process is also nearly energy independent, and weak throughout the three regions. %~\cite{BernardiGaAs}. 
In region II, the 1e1a scattering rate increases abruptly at an energy of $\omega_{\textrm{LO}}$ because the intermediate state following one LO phonon emission can be on-shell. 
The rate of the 2e processes remains negligible in region II up to an energy of $2\omega_{\textrm{LO}}$, 
above which the contribution from the 2e process increases dramatically since in region III electrons can emit two LO phonons and transition to the bottom of the conduction band.\\
\indent
The 1e1a scattering rate drops from region II to region III due to subtle reasons related to the lifetime of the intermediate state. 
An electron in region II can emit a phonon, transition to an on-shell intermediate state in region I, and then absorb another phonon to transition to a final state. 
Since the scattering rates of states in region I are considerably smaller than in the other two regions, the on-shell intermediate states have correspondingly long lifetimes, 
which gives a large amplitude to this 1e1a process. 
On the other hand, the 1e1a processes for electrons in region III lead to intermediate states in region II, which have much shorter lifetimes than in region I, 
resulting in a smaller 1e1a rate in region III compared to region II. 
% 1e1a -- explain somewhere?
We stress that including all other modes, as we have tried, does not affect these results, because the only mode that couples strongly to electrons is the LO, and the 2ph rates of the other modes are negligible compared to those of the LO mode.\\
\indent 
Let us discuss the temperature dependence of the 2ph scattering processes, focusing on the ratio $\Gamma^{(\textrm{2ph})}/\Gamma^{(\textrm{1ph})}$ of the 2ph scattering rates to the leading order 1ph rates. We provide a brief analysis here and a more extensive discussion in the Supplemental Material~\cite{SupplementalMaterial}.
The temperature dependence of the 2ph rates originates from the intermediate state lifetimes in the denominators of Eq.~(\ref{eq:lowercase_gamma_abc}) 
and the thermal factors $A^{\textrm{(i)}}$ in Eq.~(\ref{eq:thermal_factor_A}). 
Due to an increase in phonon number, the intermediate state $e$-ph scattering rate increases with temperature, 
and thus the intermediate state lifetimes become shorter, lowering the total $\Gamma^{(\textrm{2ph})}$ for increasing temperatures. 
On the contrary, the thermal factors $A^{\mathrm{(i)}}$, which contain factors proportional to the phonon number $N$, 
increase rapidly with temperature, making the total $\Gamma^{(\textrm{2ph})}$ greater at higher temperatures. 
In the 200$-$400 K temperature range investigated here, these two effects compensate,  
resulting in nearly temperature independent $\Gamma^{(\textrm{2ph})}/\Gamma^{(\textrm{1ph})}$ ratios. 
We conclude that the 2ph processes are equally as important relative to the leading order 1ph processes over a wide temperature range near room temperature.\\
%
% Figure 6
%
\begin{figure}[!t]
\includegraphics[scale=0.80]{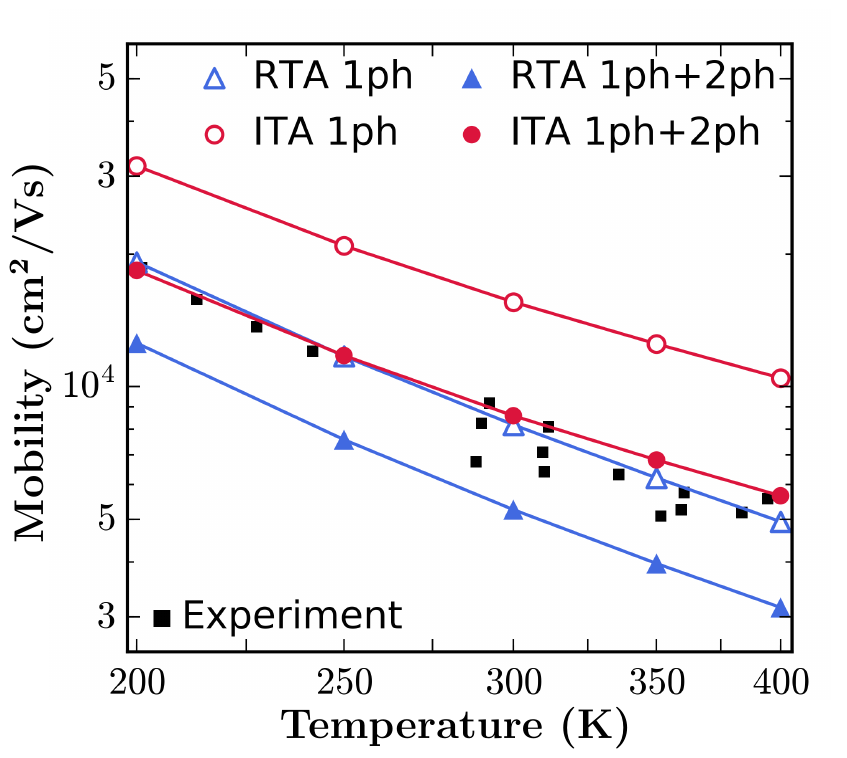}
\caption{Electron mobility in GaAs, computed by solving the linearized BTE, either within the RTA or with an iterative approach (ITA). 
For each solution method, two sets of calculations are shown, one that includes only the 1ph leading order processes, and one that includes both 1ph and 2ph scattering. 
Experimental values are taken from refs.~\cite{Rode1970, Rode1971, Blakemore1982, Hicks1969, Wolfe1970, Blood1972, Nichols1980}}
\label{fig:mobility}
\end{figure}
\indent 
Since the 2ph contributions are significant, it is natural to wonder whether they affect charge transport. 
Figure~\ref{fig:mobility} shows the electron mobility computed in GaAs by solving the BTE, either within the relaxation time approximation (RTA) 
or with a more accurate iterative approach (ITA)~\cite{Li2015, Chen2017}, which we have extended here to include 2ph processes~\cite{SupplementalMaterial}. 
Results are given for calculations that either neglect or include the 2ph contributions. 
The iterative solution with only leading order $e$-ph interactions overestimates the electron mobility in GaAs by 40$-$80\% at 200$-$400 K, 
consistent with results by other authors~\cite{Chen2017,Li2018}. 
This result is puzzling, since the BTE can accurately predict the mobility in nonpolar semiconductors; 
the discrepancy with experiment is too large to be due to small errors on the electron effective mass.\\
\indent  
This open problem is solved here by including the 2ph processes in the ITA,   
which lowers the mobility due to the additional scattering processes, giving a mobility in excellent agreement with experiments~\cite{Rode1970, Rode1971, Blakemore1982, Hicks1969, Wolfe1970, Blood1972, Nichols1980}. The agreement with experiment of the RTA with 1ph processes is thus shown to be due to error compensation. 
We conclude that the 2ph contributions are crucial to improving the accuracy of the computed electron mobility in GaAs. 
Since the LO mode coupling is strong in many polar semiconductors, we expect the 2ph processes to be significant in a wide range of materials.\\ 
\indent 
In summary, we derived analytic expressions for next-to-leading order $e$-ph scattering rates,  
and computed them in GaAs from first principles, analyzing their energy and temperature dependence. 
Our calculations show quantitatively that these 2ph contributions are substantial,  
even in a material like GaAs that is commonly thought to possess weak $e$-ph interactions. 
The formalism introduced here is general, and can be applied to other materials with weak to intermediate $e$-ph coupling and polaron states. 
Our work paves the way to studying higher order corrections in the $e$-ph interactions, and sets the stage for comparing leading plus next-to-leading order $e$-ph calculations with diagram resummation methods such as the cumulant approach. 
\begin{acknowledgments}
\vspace{10pt}
This work was supported by the Air Force Office of Scientific Research through the Young Investigator Program Grant FA9550-18-1-0280. 
J.-J. Z. and H.-Y. C. were supported by the National Science Foundation under Grant No. ACI- 1642443, which provided for code development. 
This research used resources of the National Energy Research Scientific Computing Center, a DOE Office of Science User Facility supported by the Office of Science of the U.S. Department of Energy under Contract No. DE-AC02-05CH11231.
\end{acknowledgments}
\bibliographystyle{apsrev4-1}
\bibliography{main_two_phonon}
\end{document}

\title{Supplemental Materials for\\``Next-to-Leading Order \textit{Ab Initio} Electron-Phonon Scattering''}

\author{Nien-En Lee}
\affiliation{Department of Applied Physics and Materials Science, California Institute of Technology, Pasadena, California 91125, USA}
\affiliation{Department of Physics, California Institute of Technology, Pasadena, California 91125, USA}

\author{Jin-Jian Zhou}
\affiliation{Department of Applied Physics and Materials Science, California Institute of Technology, Pasadena, California 91125, USA}

\author{Hsiao-Yi Chen}
\affiliation{Department of Applied Physics and Materials Science, California Institute of Technology, Pasadena, California 91125, USA}
\affiliation{Department of Physics, California Institute of Technology, Pasadena, California 91125, USA}

\author{Marco Bernardi}
\affiliation{Department of Applied Physics and Materials Science, California Institute of Technology, Pasadena, California 91125, USA}

\date{\today}
\pacs{}
\maketitle
\onecolumngrid
\section{Analytic Derivation of the Scattering Rates of Two-Phonon Processes}
\begin{figure}[h]
\includegraphics[scale=1.0]{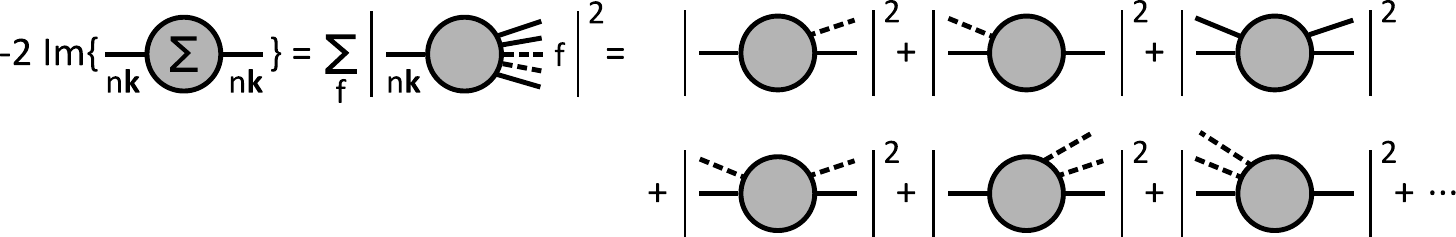}
\caption{Schematic of the scattering rate series. The first equality is the optical theorem that relates the imaginary part of the self-energy to the total scattering rates. 
Here and below, solid and dashed lines represent electron and phonon propagators, respectively. The three processes in the last row are the two-phonon processes on which we will be focusing.} 
\label{fig:scattering_schematic}
\end{figure}

In this section, we calculate the contributions to the electron-phonon ($e$-ph) scattering rates from the next-to-leading-order self-energy diagrams. We use the Matsubara technique to calculate the two-loop self-energy, whose imaginary part is related to the total scattering rates via the optical theorem (see Fig. \ref{fig:scattering_schematic}). We focus on the scattering processes with two external phonons.\\
\indent The Feynmann rules, which have been derived in Mahan \cite{Mahan2000}, will be adapted here to our context. The starting point is the $e$-ph Hamiltonian
\begin{align}
H  = \sum_{n\textbf{k}}\varepsilon_{n\textbf{k}}
      a^{\dagger}_{n\textbf{k}}a_{n\textbf{k}} 
      + \sum_{\nu\textbf{q}}\hbar\omega_{\nu\textbf{q}}
      \left(b^{\dagger}_{\nu\textbf{q}}b_{\nu\textbf{q}} 
      + \frac{1}{2}\right)
     + \frac{1}{\sqrt{N_{\Omega}}}
      \sum_{mn\textbf{k}}\sum_{\nu\textbf{q}}
      g_{mn\nu}(\textbf{k},\textbf{q})
      \left(b^{\dagger}_{\nu-\textbf{q}}+b_{\nu\textbf{q}}\right)
      a^{\dagger}_{m\textbf{k}+\textbf{q}}a_{n\textbf{k}},\nonumber
\end{align}
where $a_{n\textbf{k}}$ and $b_{\nu\textbf{q}}$ are the annihilation operators for electrons and phonons with energies $\varepsilon_{n\textbf{k}}$ and $\hbar\omega_{\nu\textbf{q}}$, respectively, $g_{mn\nu}(\textbf{k},\textbf{q})$ is the $e$-ph coupling constant, and $N_{\Omega}$ is the number of unit cells in the crystal. Comparing with Eqs. (2.67) and (3.200) in Ref. \cite{Mahan2000}, we have introduced the dependence on electron crystal momentum $\mathbf{k}$ for the $e$-ph couplings and the electron band indices $m$ and $n$. Also note that for the Hamiltonian to be Hermitian, the $e$-ph couplings must satisfy $g^{*}_{mn\nu}(\textbf{k},\textbf{q}) = g_{nm\nu}(\textbf{k}+\textbf{q},-\textbf{q})$.
%
% Figure 1
%
\vspace{-10pt}
\subsection{Feynmann Rules}
The rules for constructing diagrams are listed in Sec. 3.4 of Ref. \cite{Mahan2000}. We slightly modify them here:
\begin{enumerate}
\item With each internal electron line, associate the propagator $\mathcal{G}^{(0)}(\textbf{k},ik_{\lambda})=1/(ik_{\lambda}-\xi_{n\textbf{k}})$, where $\xi_{n\textbf{k}}=\varepsilon_{n\textbf{k}}-\mu$ and $\mu$ is the chemical potential.
\item With each internal phonon line, associate the propagator $\mathcal{D}^{(0)}(\textbf{q},i\omega_{\kappa})=-2\omega_{\nu\textbf{q}}/(\omega_{\kappa}^2+\omega_{\nu\textbf{q}}^2)$. Note that we set $\hbar=1$ for convenience here and below; it can easily be restored by dimensional analysis.
\item With each vertex, associate the $e$-ph coupling constant $g_{mn\nu}(\textbf{k},\textbf{q})$. Beware of the direction of \textbf{q}.
\item Conserve momentum and complex frequency at each vertex and sum over the internal degrees of freedom.
\item Multiply the expression by 
\begin{align}
\frac{(-1)^{L+F}(2S+1)^{F}}{(\beta N_{\Omega})^{L}}=
    \left(-\frac{1}{\beta N_{\Omega}}\right)^{L}
    \left(-2\right)^{F},\nonumber
\end{align}
where $F$ is the number of closed Fermion loops. The $(2S+1)$ factor is a summation over spin degrees of freedom, and $2S+1=2$ for electrons. The integer $L$ is the number of loops, and $\beta=1/k_{B}T$, where T is temperature.
\end{enumerate}
\begin{figure}[!h]
\includegraphics[width=0.7\textwidth]{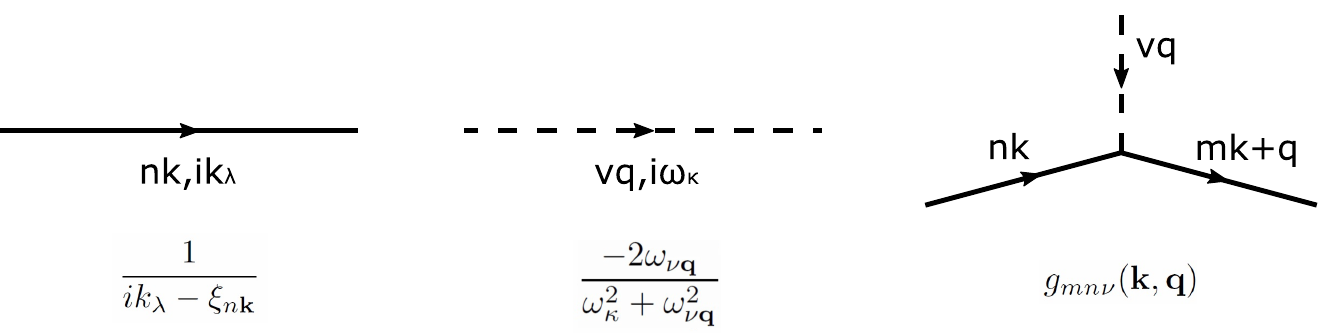}
%\caption{\label{fig:}}
\end{figure}
\subsection{Electron Self-Energy}
We consider below the 1-loop diagram that gives the lowest-order (one-phonon) self-energy and the three relevant two-loop diagrams for the electron self-energy. Diagram IIc will not contribute to the two-phonon processes and thus will not be considered in the following.
\begin{figure}[h]
\includegraphics{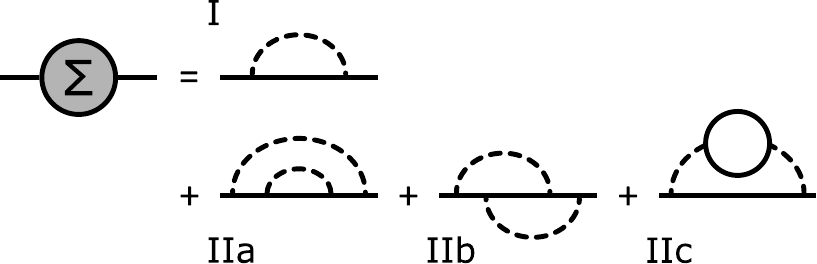}
%\caption{\label{fig:}}
\end{figure}
\vspace{15pt}
\subsection{One-Loop Diagram I}
\begin{figure}[h]
\includegraphics{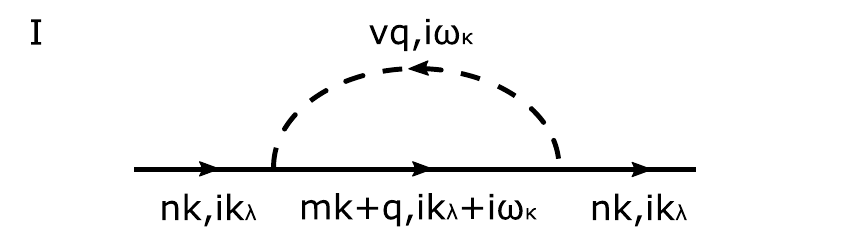}
%\caption{\label{fig:}}
\end{figure}
As a warm up exercise, we first derive the one-loop self-energy diagram, labelled as I in figure above. Since $L=1$ and $F=0$, the Feynman rules give 
\begin{align}
%\overline{\Sigma}
{\Sigma}^{\textrm{(I)}} \;& = -\frac{1}{\beta N_{\Omega}}
        \sum_{m\nu\textbf{q}} \sum_{i\omega_{\kappa}} 
        g_{mn\nu}(\textbf{k},\textbf{q})
        \mathcal{D}^{(0)}(\textbf{q},i\omega_{\kappa})
        \mathcal{G}^{(0)}
        (\textbf{k}+\textbf{q},ik_{\lambda}+i\omega_{\kappa})
        g_{nm\nu}(\textbf{k}+\textbf{q},-\textbf{q})
        \nonumber\\
    &= -\frac{1}{\beta N_{\Omega}}
        \sum_{m\nu\textbf{q}} \sum_{i\omega_{\kappa}}
        |g_{mn\nu}(\textbf{k},\textbf{q})|^{2}
        \frac{-2\omega_{\nu\textbf{q}}}
        {\omega_{\kappa}^2+\omega_{\nu\textbf{q}}^2}
        \frac{1}{ik_{\lambda}+i\omega_{\kappa}-
        \xi_{m\textbf{k}+\textbf{q}}}
        \nonumber\\
    &\equiv \frac{1}{N_{\Omega}}
        \sum_{m\nu\textbf{q}} 
        |g_{mn\nu}(\textbf{k},\textbf{q})|^{2}
        \left( -\frac{1}{\beta} \right)
        \sum_{i\omega_{\kappa}}f(i\omega_{\kappa})
        \nonumber,
\end{align}
where $f(z)$ is defined as
\begin{align}
f(z) \equiv \frac{2\omega_{\nu\textbf{q}}}
            {z^{2}-\omega_{\nu\textbf{q}}^{2}}\;
            \frac{1}{z+ik_
            {\lambda}-\xi_{m\textbf{k}+\textbf{q}}}.
            \nonumber
\end{align}
\indent To apply the Matsubara frequency summation method, we define the bosonic weighting function as in \cite{Mahan2000}:
\begin{align}
n_{B}(z)=\frac{1}{e^{\beta z}-1},\nonumber
\end{align}
whose poles are at $i\omega_{\kappa}=i2\kappa \pi/\beta$, with residues $1/\beta$ and integer $\kappa$ values. The weighting function for fermions is
\begin{align}
n_{F}(z)=\frac{1}{e^{\beta z}+1},\nonumber
\end{align}
whose poles are at $i k_{\lambda}=i(2\lambda +1) \pi /\beta$, with residues $-1/\beta$ and integer $\lambda$ values.\\
\indent For this diagram, we will do the contour integral for $f(z)n_{B}(z)$ at the complex infinity. Since $f(z)n_{B}(z)$ decays faster than $1/z$, we can apply the Cachy residue theorem, which gives (here and below, $z'$ are the relevant poles):
\begin{align}
0 =&\; \lim_{|z|\to\infty} \oint \frac{dz}{2\pi i}f(z)n_{B}(z)
        \nonumber\\
    =&\; \sum_{z^{\prime} \textrm{ of } f\cdot n_{B}}
        \textrm{Res}\{ f(z^{\prime})n_{B}(z^{\prime}) \}
        \nonumber\\
    =&\; \sum_{i\omega_{\kappa}}f(i\omega_{\kappa})\frac{1}{\beta}
        + \sum_{z^{\prime} \textrm{ of } f}
        \textrm{Res}\{ f(z^{\prime}) n_{B}(z^{\prime}) \}
        \nonumber.
\end{align}
Using this result, we get:
\begin{align}
%\overline
{\Sigma}^{\textrm{(I)}} = \frac{1}{N_{\Omega}}
       \sum_{m\nu\textbf{q}} 
       |g_{mn\nu}(\textbf{k},\textbf{q})|^{2}
       \sum_{z^{\prime} \textrm{ of } f}
       \textrm{Res}\{ f(z^{\prime}) \} n_{B}(z^{\prime})
       \nonumber.
\end{align}
The poles of $f(z)$ are at $z_{1}=\omega_{\nu\textbf{q}}$, $z_{2}=-\omega_{\nu\textbf{q}}$ and $z_{3}=-ik_{\lambda}+\xi_{m\textbf{k}+\textbf{q}}$. Their residues are
\begin{align}
\textrm{Res}\{ f(z), z_{1} \} &\;= \frac{1}{\omega_{\nu\textbf{q}}+
        ik_{\lambda}-\xi_{m\textbf{k}+\textbf{q}}}
        \nonumber\\
\textrm{Res}\{ f(z), z_{2} \} &\;= -\frac{1}{-\omega_{\nu\textbf{q}}+
        ik_{\lambda}-\xi_{m\textbf{k}+\textbf{q}}}
        \nonumber\\
\textrm{Res}\{ f(z), z_{3} \} &\;= \frac{2\omega_{\nu\textbf{q}}}
        {(-ik_{\lambda}+\xi_{m\textbf{k}+\textbf{q}})^{2}
        -\omega_{\nu\textbf{q}}^{2}}
      =\frac{1}{-\omega_{\nu\textbf{q}}+
        ik_{\lambda}-\xi_{m\textbf{k}+\textbf{q}}}
        -\frac{1}{\omega_{\nu\textbf{q}}+
        ik_{\lambda}-\xi_{m\textbf{k}+\textbf{q}}}
        .\nonumber
\end{align}
We also know that $n_{B}(z_{1})=n_{B}(\omega_{\nu\textbf{q}}) \equiv N_{\nu\textbf{q}}$, $n_{B}(z_{2})=n_{B}(-\omega_{\nu\textbf{q}})=-N_{\nu\textbf{q}}-1$ and $n_{B}(z_{3})= -n_{F}(\xi_{m\textbf{k}+\textbf{q}}) \equiv -f_{m\textbf{k}+\textbf{q}}$, where $N$ and $f$ are the thermal occupation numbers for phonons and electrons, respectively. We also used the fact that $ik_{\lambda}=i(2\lambda +1) \pi /\beta$. Substituting this result in the self-energy expression, we get
\begin{align}
%\overline
{\Sigma}^{\textrm{(I)}} = \frac{1}{N_{\Omega}}
       \sum_{m\nu\textbf{q}}
       |g_{mn\nu}(\textbf{k},\textbf{q})|^{2}
       \left[
       \frac{N_{\nu\textbf{q}}+f_{m\textbf{k}+\textbf{q}}}
       {ik_{\lambda}+\omega_{\nu\textbf{q}}
        -\xi_{m\textbf{k}+\textbf{q}}}
       +\frac{1+N_{\nu\textbf{q}}-f_{m\textbf{k}+\textbf{q}}}
       {ik_{\lambda}-\omega_{\nu\textbf{q}}
        -\xi_{m\textbf{k}+\textbf{q}}}
       \right]
       \nonumber.
\end{align}
Employing the analytic continuation $ik_{\lambda}\rightarrow E+i\eta$, we obtain the off-shell lowest-order $e$-ph self-energy:
% \{see Eq. (3.140) in Ref. \cite{Mahan2000}\}
\begin{align}
\Sigma^{\textrm{(I)}}(E) = \frac{1}{N_{\Omega}}
       \sum_{m\nu\textbf{q}}
       |g_{mn\nu}(\textbf{k},\textbf{q})|^{2}
       \left[
       \frac{N_{\nu\textbf{q}}+f_{m\textbf{k}+\textbf{q}}}
       {E+\omega_{\nu\textbf{q}}
        -\xi_{m\textbf{k}+\textbf{q}}+i\eta}
       +\frac{1+N_{\nu\textbf{q}}-f_{m\textbf{k}+\textbf{q}}}
       {E-\omega_{\nu\textbf{q}}
        -\xi_{m\textbf{k}+\textbf{q}}+i\eta}
       \right]
       \nonumber.
\end{align}
We will be mainly interested in the scattering rate at the electron energy $\xi_{n\textbf{k}}$ and therefore we will set $E=\xi_{n\textbf{k}}$ to obtain the on-shell self-energy for the state with band $n$ and crystal momentum \textbf{k}. Using the identity
\begin{align}
\frac{1}{x+i\eta}=P{\frac{1}{x}}-i\pi \delta(x) \nonumber
\end{align}
and Eq. (7.304) in Ref. \cite{Mahan2000}, which states that the scattering rate $\Gamma$ is obtained as $\Gamma=-(2/\hbar)\textrm{Im}\Sigma$, we get
\begin{align}
\Gamma^{\textrm{(I)}}_{n\textbf{k}} =
       \frac{2\pi}{\hbar} \frac{1}{N_{\Omega}}
       \sum_{m\nu\textbf{q}}
       |g_{mn\nu}(\textbf{k},\textbf{q})|^{2}
    \left[
       (N_{\nu\textbf{q}}+f_{m\textbf{k}+\textbf{q}})
       \delta(\xi_{n\textbf{k}}+\hbar\omega_{\nu\textbf{q}}
        -\xi_{m\textbf{k}+\textbf{q}})
       + \{(1+N_{\nu\textbf{q}}-f_{m\textbf{k}+\textbf{q}})
       \delta(\xi_{n\textbf{k}}-\hbar\omega_{\nu\textbf{q}}
        -\xi_{m\textbf{k}+\textbf{q}})
        \right],
        \nonumber
\end{align}
with $\hbar$ placed back into the expression. This is the well-known lowest-order scattering rate commonly used in first-principles calculations.
\subsection{Two-Loop Diagram IIa}
Now we compute the first two-loop diagram, which is shown in the figure below. Since $L=2$ and $F=0$, the Feynman rules give
\begin{figure}[h]
\includegraphics{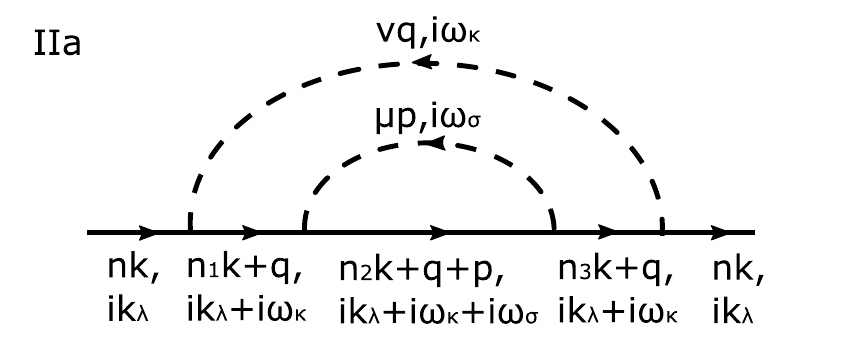}
%\caption{\label{fig:}}
\end{figure}
\begin{align}
%\overline
{\Sigma}^{\textrm{(IIa)}} = \frac{1}{\beta^{2} N_{\Omega}^{2}}
        \sum_{n_{1}n_{2}n_{3}}
        \sum_{\nu\textbf{q}}
        \sum_{\mu\textbf{p}}
        g_{n_{1}n\nu}(\textbf{k},\textbf{q})
        g^{*}_{n_{3}n\nu}(\textbf{k},\textbf{q})
        g_{n_{2}n_{1}\mu}(\textbf{k}+\textbf{q},\textbf{p})
        g^{*}_{n_{2}n_{3}\mu}(\textbf{k}+\textbf{q},\textbf{p})
        \sum_{i\omega_{\kappa}}
        \sum_{i\omega_{\sigma}}
        f(i\omega_{\kappa},i\omega_{\sigma})
        \nonumber,
\end{align}
where $f(i\omega_{\kappa},i\omega_{\sigma})$ is defined as
\begin{align}
f(i\omega_{\kappa},i\omega_{\sigma}) &\;
   \equiv 
        \frac{1}{ik_{\lambda}+i\omega_{\kappa}
        -\xi_{n_{1}\textbf{k}+\textbf{q}}}\;
        \frac{1}{ik_{\lambda}+i\omega_{\kappa}+i\omega_{\sigma}
        -\xi_{n_{2}\textbf{k}+\textbf{q}+\textbf{p}}} \;
        \frac{1}{ik_{\lambda}+i\omega_{\kappa}
        -\xi_{n_{3}\textbf{k}+\textbf{q}}}\;
        \frac{2\omega_{\nu\textbf{q}}}
        {\omega_{\kappa}^{2}+\omega_{\nu\textbf{q}}^{2}}\;
        \frac{2\omega_{\mu\textbf{p}}}
        {\omega_{\sigma}^{2}+\omega_{\mu\textbf{p}}^{2}}
        \nonumber\\[3ex]
   &\equiv  A_{\kappa} \;
        \frac{1}{ik_{\lambda}+i\omega_{\kappa}+i\omega_{\sigma}
        -\xi_{n_{2}\textbf{k}+\textbf{q}+\textbf{p}}} \;
        \frac{2\omega_{\mu\textbf{p}}}
        {-(i\omega_{\sigma})^{2}+\omega_{\mu\textbf{p}}^{2}}
        \nonumber,
\end{align}
where in $A_{\kappa}$ we collect all terms independent of $i\omega_{\sigma}$. Let us sum over $i\omega_{\sigma}$ first. Performing the contour integral for $f(i\omega_{\kappa},z)n_{B}(z)$ gives
\begin{align}
\sum_{i\omega_{\sigma}}
        f(i\omega_{\kappa},i\omega_{\sigma})
    = -\beta
        \sum_{z^{\prime} \textrm{ of } f}
        \textrm{Res}\{ 
        f(i\omega_{\kappa},z^{\prime}) 
        n_{B}(z^{\prime}) \}
        \nonumber.
\end{align}
Note that $f(i\omega_{\kappa},z)$ has three poles, whose residues and bosonic weight factors are computed as:
\begin{align}
z_{1} &\; \rightarrow -ik_{\lambda}-i\omega_{\kappa}
        +\xi_{n_{2}\textbf{k}+\textbf{q}+\textbf{p}}
        \nonumber\\
\textrm{Res}\{ f(i\omega_{\kappa},z_{1}) \}
        &\; =A_{\kappa} 
        \left( 
        \frac{1}{z_{1}+\omega_{\mu\textbf{p}}}
        -\frac{1}{z_{1}-\omega_{\mu\textbf{p}}}
        \right)\nonumber\\
n_{B}(z_{1}) &\;= -f_{n_{2}\textbf{k}+\textbf{q}+\textbf{p}}
        \nonumber\\[4ex]
z_{2} &\; \rightarrow \omega_{\mu\textbf{p}}
        \nonumber\\
\textrm{Res}\{ f(i\omega_{\kappa},z_{2}) \}
        &\; = -A_{\kappa}
        \frac{1}
        {z_{2}+ik_{\lambda}+i\omega_{\kappa}
        -\xi_{n_{2}\textbf{k}+\textbf{q}+\textbf{p}}}
        \nonumber\\
n_{B}(z_{2}) &\;= N_{\mu\textbf{p}}
        \nonumber\\[4ex]
z_{3} &\; \rightarrow -\omega_{\mu\textbf{p}}
        \nonumber\\
\textrm{Res}\{ f(i\omega_{\kappa},z_{3}) \}
        &\; = A_{\kappa}
        \frac{1}
        {z_{3}+ik_{\lambda}+i\omega_{\kappa}
        -\xi_{n_{2}\textbf{k}+\textbf{q}+\textbf{p}}}
        \nonumber\\
n_{B}(z_{3}) &\;= -N_{\mu\textbf{p}}-1
        \nonumber
\end{align}
Using these results, we get
\begin{align}
\sum_{i\omega_{\sigma}}
        f(i\omega_{\kappa},i\omega_{\sigma})
    = \beta A_{\kappa}
        \frac{1+N_{\mu\textbf{p}}
        -f_{n_{2}\textbf{k}+\textbf{q}+\textbf{p}}}
        {ik_{\lambda}+i\omega_{\kappa}
        -\xi_{n_{2}\textbf{k}+\textbf{q}+\textbf{p}}
        -\omega_{\mu\textbf{p}}}
    + \beta A_{\kappa}
        \frac{N_{\mu\textbf{p}}
        +f_{n_{2}\textbf{k}+\textbf{q}+\textbf{p}}}
        {ik_{\lambda}+i\omega_{\kappa}
        -\xi_{n_{2}\textbf{k}+\textbf{q}+\textbf{p}}
        +\omega_{\mu\textbf{p}}}
    \equiv \beta h^{(-)}(i\omega_{\kappa})
        + \beta h^{(+)}(i\omega_{\kappa})
        \nonumber,
\end{align}
where we defined two functions, $h^{(-)}(i\omega_{\kappa})$ and $h^{(+)}(i\omega_{\kappa})$, as the first and second terms in the expression above.\\
\indent
We then sum over $i\omega_{\kappa}$, using again
\begin{align}
\sum_{i\omega_{\kappa}}
        h^{(\pm)}(i\omega_{\kappa})
    = -\beta
        \sum_{z^{\prime} \textrm{ of } h^{(\pm)}}
        \textrm{Res}\{ 
        h^{(\pm)}(z^{\prime}) 
        n_{B}(z^{\prime}) \}
        \nonumber.
\end{align}
A subtle point is that the cases with $n_{3} \neq n_{1}$ and $n_{3} = n_{1}$ have different pole structures, and need to be discussed separately (see the figure above for diagram IIa; $n_3$ and $n_1$ label two intermediate electronic states in the self-energy diagram). Luckily, the two cases give the same expression for the two-phonon scattering processes, as we show explicitly below. Before carrying out the calculation, let us introduce some useful abbreviations. We will use in the following $\xi_{n_{1}\textbf{k}+\textbf{q}} \equiv \xi_{1}$, $\xi_{n_{1}\textbf{k}+\textbf{p}} \equiv \xi_{1\textbf{p}}$, $f_{n_{2}\textbf{k}+\textbf{q}+\textbf{p}} \equiv f_{2}$, $\omega_{\nu\textbf{q}} \equiv \omega_{\textbf{q}}$, etc.
%
%
%
%\newpage
%\onecolumngrid
%\vspace{1cm}
\subsubsection{Case with $n_{3} \neq n_{1}$}
Let us focus on $h^{(-)}$ for the case $n_{3} \neq n_{1}$ first. In this case, $h^{(-)}$ is defined as
\begin{align}
h^{(-)}(z)= \nonumber
 \frac{1}{z+ik_{\lambda}-\xi_{1}} \;
        \frac{1}{z+ik_{\lambda}-\xi_{3}} \;
        \frac{-2\omega_{\textbf{q}}}
            {z^{2}-\omega_{\textbf{q}}^{2}} \;
        \frac{1+N_{\textbf{p}}-f_{2}}
            {z+ik_{\lambda}-\xi_{2}-\omega_{\textbf{p}}}
        \nonumber.
\end{align}
It has five poles, which are given here together with their residues and bosonic weight factors:
\begin{align}
z_{1} &\; \rightarrow -ik_{\lambda}+\xi_{1}
        \nonumber\\
\textrm{Res}\{ h^{(-)}(z_{1}) \} &\; =
        \frac{1}{\xi_{1}-\xi_{3}}\;
        \frac{1+N_{\textbf{p}}-f_{2}}
            {\xi_{1}-\xi_{2}-\omega_{\textbf{p}}}
        \left( 
        \frac{1}{-ik_{\lambda}+\xi_{1}+\omega_{\textbf{q}}}
        -\frac{1}{-ik_{\lambda}+\xi_{1}-\omega_{\textbf{q}}}
        \right)\nonumber\\
n_{B}(z_{1}) &\;= -f_{1}
        \nonumber\\[4ex]
z_{2} &\; \rightarrow -ik_{\lambda}+\xi_{3}
        \nonumber\\
\textrm{Res}\{ h^{(-)}(z_{2}) \} &\; =
        \frac{1}{\xi_{3}-\xi_{1}}\;
        \frac{1+N_{\textbf{p}}-f_{2}}
            {\xi_{3}-\xi_{2}-\omega_{\textbf{p}}}
        \left( 
        \frac{1}{-ik_{\lambda}+\xi_{3}+\omega_{\textbf{q}}}
        -\frac{1}{-ik_{\lambda}+\xi_{3}-\omega_{\textbf{q}}}
        \right)\nonumber\\
n_{B}(z_{2}) &\;= -f_{3}
        \nonumber\\[4ex]
z_{3} &\; \rightarrow \omega_{\textbf{q}}
        \nonumber\\
\textrm{Res}\{ h^{(-)}(z_{3}) \}
        &\; = 
        \frac{1}{ik_{\lambda}+\omega_{\textbf{q}}-\xi_{1}}\;
        \frac{1}{ik_{\lambda}+\omega_{\textbf{q}}-\xi_{3}}
            \;\left(-1\right)\;
            \frac{1+N_{\textbf{p}}-f_{2}}
            {ik_{\lambda}+\omega_{\textbf{q}}
                -\xi_{2}-\omega_{\textbf{p}}}
            \nonumber\\
n_{B}(z_{3}) &\;= N_{\textbf{q}}
        \nonumber\\[4ex]
z_{4} &\; \rightarrow -\omega_{\textbf{q}}
        \nonumber\\
\textrm{Res}\{ h^{(-)}(z_{4}) \}
        &\; = 
        \frac{1}{ik_{\lambda}-\omega_{\textbf{q}}-\xi_{1}}\;
        \frac{1}{ik_{\lambda}-\omega_{\textbf{q}}-\xi_{3}}\;
            \frac{1+N_{\textbf{p}}-f_{2}}
            {ik_{\lambda}-\omega_{\textbf{q}}
                -\xi_{2}-\omega_{\textbf{p}}}
            \nonumber\\
n_{B}(z_{4}) &\;= -N_{\textbf{q}}-1
        \nonumber\\[4ex]
z_{5} &\; \rightarrow -ik_{\lambda}+\xi_{2}+\omega_{\textbf{p}}
        \nonumber\\
\textrm{Res}\{ h^{(-)}(z_{5}) \} &\; =
        \frac{1}{\xi_{2}+\omega_{\textbf{p}}-\xi_{1}}\;
            \frac{1}{\xi_{2}+\omega_{\textbf{p}}-\xi_{3}}
            \left( 
            \frac{1}
            {-ik_{\lambda}+\xi_{2}
                +\omega_{\textbf{p}}+\omega_{\textbf{q}}}
            -\frac{1}
            {-ik_{\lambda}+\xi_{2}
                +\omega_{\textbf{p}}-\omega_{\textbf{q}}}
        \right)\left(1+N_{\textbf{p}}-f_2\right)
        \nonumber\\
n_{B}(z_{5}) &\;= 
        -\frac{N_{\textbf{p}}f_{2}}{1+N_{\textbf{p}}-f_{2}}
        \nonumber
\end{align}
The terms related to the two-phonon processes are those containing $1/(ik- \xi_{2} \pm \omega_{\bf{p}} \pm \omega_{\bf{q}})$. After repeating this procedure for $h^{(+)}$, collecting terms and ignoring terms that are irrelevant to the two-phonon processes, we get
\begin{align}
\frac{1}{\beta^2} \sum_{i\omega_{\kappa}}\sum_{i\omega_{\sigma}}
        f(i\omega_{\kappa},i\omega_{\sigma}) \;=\; &
    \frac{1}{ik_{\lambda}- \xi_{2}- \omega_{\bf{p}}+ \omega_{\bf{q}}}
    \left[\;
        \frac{1+ N_{\bf{p}}- f_2}
        {ik_{\lambda}-\xi_1+\omega_{\bf{q}}}\;
        \frac{N_{\bf{q}}}{ik_{\lambda}-\xi_3+\omega_{\bf{q}}}
       +\frac{N_{\bf{p}}f_2}{\xi_2-\xi_1+\omega_{\bf{p}}}\;
        \frac{1}{\xi_2-\xi_3+\omega_{\bf{p}}}
    \;\right]\nonumber\\
&+
    \frac{1}{ik_{\lambda}- \xi_{2}- \omega_{\bf{p}}- \omega_{\bf{q}}}
    \left[\;
        \frac{1+ N_{\bf{p}}- f_2}
        {ik_{\lambda}-\xi_1-\omega_{\bf{q}}}\;
        \frac{1+N_{\bf{q}}}{ik_{\lambda}-\xi_3-\omega_{\bf{q}}}
       -\frac{N_{\bf{p}}f_2}{\xi_2-\xi_1+\omega_{\bf{p}}}\;
        \frac{1}{\xi_2-\xi_3+\omega_{\bf{p}}}
    \;\right]\nonumber\\
&+
    \frac{1}{ik_{\lambda}- \xi_{2}+ \omega_{\bf{p}}+ \omega_{\bf{q}}}
    \left[\;
        \frac{N_{\bf{p}}+ f_2}{ik_{\lambda}-\xi_1+\omega_{\bf{q}}}\;
        \frac{N_{\bf{q}}}{ik_{\lambda}-\xi_3+\omega_{\bf{q}}}
       +\frac{f_2\;(1+N_{\bf{p}})}{\xi_2-\xi_1-\omega_{\bf{p}}}\;
        \frac{1}{\xi_2-\xi_3-\omega_{\bf{p}}}
    \;\right]\nonumber\\
&+
    \frac{1}{ik_{\lambda}- \xi_{2}+ \omega_{\bf{p}}- \omega_{\bf{q}}}
    \left[\;
        \frac{N_{\bf{p}}+ f_2}{ik_{\lambda}-\xi_1-\omega_{\bf{q}}}\;
        \frac{1+N_{\bf{q}}}{ik_{\lambda}-\xi_3-\omega_{\bf{q}}}
       -\frac{f_2\;(1+N_{\bf{p}})}{\xi_2-\xi_1-\omega_{\bf{p}}}\;
        \frac{1}{\xi_2-\xi_3-\omega_{\bf{p}}}
    \;\right]\nonumber\\
&+\cdots\nonumber
\end{align}
The rates of the two-phonon processes emerge after analytically continuing $ik_{\lambda}$ to $E+i\eta$ and taking the imaginary part of $1/(E- \xi_{2} \pm \omega_{\bf{p}} \pm \omega_{\bf{q}}+ i\eta)$. We also use the delta functions to set $E=\xi_2 \mp \omega_{\bf{p}} \mp \omega_{\bf{q}}$ in some of the denominators. After carrying out these calculations, we obtain
\begin{align}
\textrm{Im}
\left\{
\frac{1}{\beta^2} \sum_{i\omega_{\kappa}}\sum_{i\omega_{\sigma}}
        f(i\omega_{\kappa},i\omega_{\sigma})\right\}
        \;=\; &
    -i\pi\delta(E- \xi_{2}- \omega_{\bf{p}}+ \omega_{\bf{q}})
        \frac{1}{\xi_2-\xi_1+\omega_{\bf{p}}}\;
        \frac{1}{\xi_2-\xi_3+\omega_{\bf{p}}}
    \left[\;
        (1+N_{\bf{p}}-f_2)N_{\bf{q}}+N_{\bf{p}}f_2
    \;\right]\nonumber\\
&
    -i\pi\delta(E- \xi_{2}- \omega_{\bf{p}}- \omega_{\bf{q}})
        \frac{1}{\xi_2-\xi_1+\omega_{\bf{p}}}\;
        \frac{1}{\xi_2-\xi_3+\omega_{\bf{p}}}
    \left[\;
        (1+N_{\bf{p}}-f_2)(1+N_{\bf{q}})-N_{\bf{p}}f_2
    \;\right]\nonumber\\
&
    -i\pi\delta(E- \xi_{2}+ \omega_{\bf{p}}+ \omega_{\bf{q}})
        \frac{1}{\xi_2-\xi_1-\omega_{\bf{p}}}\;
        \frac{1}{\xi_2-\xi_3-\omega_{\bf{p}}}
    \left[\;
        (N_{\bf{p}}+f_2)N_{\bf{q}}+f_2(1+N_{\bf{p}})
    \;\right]\nonumber\\
&
    -i\pi\delta(E- \xi_{2}+ \omega_{\bf{p}}- \omega_{\bf{q}})
        \frac{1}{\xi_2-\xi_1-\omega_{\bf{p}}}\;
        \frac{1}{\xi_2-\xi_3-\omega_{\bf{p}}}
    \left[\;
        (N_{\bf{p}}+f_2)(1+N_{\bf{q}})-f_2(1+N_{\bf{p}})
    \;\right]\nonumber\\
&+\cdots\nonumber
\end{align}
%
%\newpage
%
%
%
\subsubsection{Case with $n_{3} = n_{1}$}
In this case, $h^{(-)}$ is defined as
\begin{align}
h^{(-)}(z)= 
 \left(\frac{1}{z+ik_{\lambda}-\xi_{1}} \right)^{2} \;
        \frac{-2\omega_{\textbf{q}}}
            {z^{2}-\omega_{\textbf{q}}^{2}} \;
        \frac{1+N_{\textbf{p}}-f_{2}}
            {z+ik_{\lambda}-\xi_{2}-\omega_{\textbf{p}}}
        \nonumber.
\end{align}
The function $h(z)n_{B}(z)$ has a pole of order $2$ at $z_{1}=-ik_{\lambda}+\xi_{1}$. By employing
\begin{align}
\textrm{Res}\{ f, z_{1} \} = \frac{1}{(n-1)!}
    \lim_{z\rightarrow z_{1}} \frac{d^{n-1}}{dz^{n-1}}
    \{ (z-z_{1})^{n}f(z) \}\nonumber,
\end{align}
where $n$ is the order of the pole, we get
\begin{align}
\textrm{Res}\{ h^{(-)}n_{B}, z_{1} \} =\; &
        \frac{4z_{1}\omega_{\textbf{q}}}
            {(z_{1}^{2}-\omega_{\textbf{q}}^{2})^{2}}\;
        \frac{1+N_{\textbf{p}}-f_{2}}
            {z_{1}+ik_{\lambda}-\xi_{2}-\omega_{\textbf{p}}}\;
        n_{B}(z_{1})
    + \frac{2\omega_{\textbf{q}}}
            {z_{1}^{2}-\omega_{\textbf{q}}^{2}}\;
        \frac{1+N_{\textbf{p}}-f_{2}}
            {(z_{1}+ik_{\lambda}-\xi_{2}-\omega_{\textbf{p}})^{2}}\;
        n_{B}(z_{1})\nonumber\\[3ex]
    & + \frac{-2\omega_{\textbf{q}}}
            {z_{1}^{2}-\omega_{\textbf{q}}^{2}}\;
        \frac{1+N_{\textbf{p}}-f_{2}}
            {z_{1}+ik_{\lambda}-\xi_{2}-\omega_{\textbf{p}}}\;
        n_{B}^{\prime}(z_{1}).
        \nonumber
\end{align}
After substituting $z_{1}=-ik_{\lambda}+\xi_{1}$, $n_{B}(z_{1})=-f_{1}$ and $n_{B}^{\prime}(z_{1})=\beta f_{1}(1-f_{1})$, we get
\begin{align}
\textrm{Res}\{ h^{(-)}n_{B}, z_{1} \} = \; &
    \frac{f_{1}(1+N_{\textbf{p}}-f_{2})}
        {(\xi_{1}-\xi_{2}-\omega_{\textbf{p}})^{2}}\;
        \frac{-2\omega_{\textbf{q}}}
        {(ik_{\lambda}-\xi_{1})^{2}-\omega_{\textbf{q}}^{2}}
    \;+\; \frac{\beta f_{1}(1-f_{1})(1+N_{\textbf{p}}-f_{2})}
        {\xi_{1}-\xi_{2}-\omega_{\textbf{p}}}\;
        \frac{-2\omega_{\textbf{q}}}
        {(ik_{\lambda}-\xi_{1})^{2}-\omega_{\textbf{q}}^{2}}\;
        \nonumber\\[3ex]
    & -\; \frac{f_{1}(1+N_{\textbf{p}}-f_{2})}
        {\xi_{1}-\xi_{2}-\omega_{\textbf{p}}}
        \left[
        \frac{1}
        {(ik_{\lambda}-\xi_{1}+\omega_{\textbf{q}})^{2}}
        -\frac{1}
        {(ik_{\lambda}-\xi_{1}-\omega_{\textbf{q}})^{2}}
        \right]
        \nonumber.
\end{align}
The other three poles are simple poles and can be treated in the usual way. Repeating this procedure for $h^{(+)}$ and adding all the contributions, we get
\begin{align}
\frac{1}{\beta^{2}} \sum_{i\omega_{\kappa}}
        \sum_{i\omega_{\sigma}} \;&
        f(i\omega_{\kappa},i\omega_{\sigma}) =\nonumber\\
    &\frac{2\omega_{\textbf{q}}(N_{\textbf{p}}+f_{2})}
        {(ik_{\lambda}-\xi_{1})^{2}-\omega_{\textbf{q}}^{2}}
        \left[
        \frac{f_{1}}{(\xi_{1}-\xi_{2}+\omega_{\textbf{p}})^{2}}
       +\frac{\beta f_{1}(1-f_{1})}
        {\xi_{1}-\xi_{2}+\omega_{\textbf{p}}}
        \right]
    +\;\frac{2\omega_{\textbf{q}}(1+N_{\textbf{p}}-f_{2})}
        {(ik_{\lambda}-\xi_{1})^{2}-\omega_{\textbf{q}}^{2}}
        \left[
        \frac{f_{1}}{(\xi_{1}-\xi_{2}-\omega_{\textbf{p}})^{2}}
       +\frac{\beta f_{1}(1-f_{1})}
        {\xi_{1}-\xi_{2}-\omega_{\textbf{p}}}
        \right]
        \nonumber\\[3ex]
    &-\;\frac{2\omega_{\textbf{q}}f_{2}(1+N_{\textbf{p}})}
            {(ik_{\lambda}-\xi_{2}+\omega_{\textbf{p}})^{2}
            -\omega_{\textbf{q}}^{2}}
        \left(
        \frac{1}{\xi_{1}-\xi_{2}+\omega_{\textbf{p}}}
        \right)^{2}
    -\;\frac{2\omega_{\textbf{q}}f_{2}N_{\textbf{p}}}
            {(ik_{\lambda}-\xi_{2}-\omega_{\textbf{p}})^{2}
            -\omega_{\textbf{q}}^{2}}
        \left(
        \frac{1}{\xi_{1}-\xi_{2}-\omega_{\textbf{p}}}
        \right)^{2}
        \nonumber\\[3ex]
    &+\;(N_{\textbf{p}}+f_{2})\left(
        \frac{1}{ik_{\lambda}-\xi_{1}+\omega_{\textbf{q}}}
        \right)^{2}
    \; \left[
        \frac{f_{1}}{\xi_{1}-\xi_{2}+\omega_{\textbf{p}}}
        +\frac{N_{\textbf{q}}}
        {ik_{\lambda}-\xi_{2}+\omega_{\textbf{p}}+\omega_{\textbf{q}}}
        \right]
        \nonumber\\[3ex]
    &+\;(1+N_{\textbf{p}}-f_{2})\left(
        \frac{1}{ik_{\lambda}-\xi_{1}+\omega_{\textbf{q}}}
        \right)^{2}
    \; \left[
        \frac{f_{1}}{\xi_{1}-\xi_{2}-\omega_{\textbf{p}}}
        +\frac{N_{\textbf{q}}}
        {ik_{\lambda}-\xi_{2}-\omega_{\textbf{p}}+\omega_{\textbf{q}}}
        \right]
        \nonumber\\[3ex]
    &+\;(N_{\textbf{p}}+f_{2})\left(
        \frac{1}{ik_{\lambda}-\xi_{1}-\omega_{\textbf{q}}}
        \right)^{2}
    \; \left[
        \frac{-f_{1}}{\xi_{1}-\xi_{2}+\omega_{\textbf{p}}}
        +\frac{1+N_{\textbf{q}}}
        {ik_{\lambda}-\xi_{2}+\omega_{\textbf{p}}-\omega_{\textbf{q}}}
        \right]
        \nonumber\\[3ex]
    &+\;(1+N_{\textbf{p}}-f_{2})\left(
        \frac{1}{ik_{\lambda}-\xi_{1}-\omega_{\textbf{q}}}
        \right)^{2}
    \;  \left[
        \frac{-f_{1}}{\xi_{1}-\xi_{2}-\omega_{\textbf{p}}}
        +\frac{1+N_{\textbf{q}}}
        {ik_{\lambda}-\xi_{2}-\omega_{\textbf{p}}-\omega_{\textbf{q}}}
        \right]
        \nonumber\\[4ex]
=\;\;&\frac{1}{ik_{\lambda}- \xi_{2}- \omega_{\bf{p}}+ \omega_{\bf{q}}}
    \left[\;
        \frac{N_{\bf{q}}(1+ N_{\bf{p}}- f_2)}
        {(ik_{\lambda}-\xi_1+\omega_{\bf{q}})^2}
       +\frac{N_{\bf{p}}f_2}{(\xi_2-\xi_1+\omega_{\bf{p}})^2}
    \;\right]\nonumber\\[3ex]
&+
    \frac{1}{ik_{\lambda}- \xi_{2}- \omega_{\bf{p}}- \omega_{\bf{q}}}
    \left[\;
        \frac{(1+N_{\bf{q}})(1+ N_{\bf{p}}- f_2)}
        {(ik_{\lambda}-\xi_1-\omega_{\bf{q}})^2}
       -\frac{N_{\bf{p}}f_2}{(\xi_2-\xi_1+\omega_{\bf{p}})^2}
    \;\right]\nonumber\\[3ex]
&+
    \frac{1}{ik_{\lambda}- \xi_{2}+ \omega_{\bf{p}}+ \omega_{\bf{q}}}
    \left[\;
        \frac{N_{\bf{q}}(N_{\bf{p}}+ f_2)}
        {(ik_{\lambda}-\xi_1+\omega_{\bf{q}})^2}
       +\frac{f_2\;(1+N_{\bf{p}})}{(\xi_2-\xi_1-\omega_{\bf{p}})^2}
    \;\right]\nonumber\\[3ex]
&+
    \frac{1}{ik_{\lambda}- \xi_{2}+ \omega_{\bf{p}}- \omega_{\bf{q}}}
    \left[\;
        \frac{(1+N_{\bf{q}})(N_{\bf{p}}+ f_2)}
        {(ik_{\lambda}-\xi_1-\omega_{\bf{q}})^2}
       -\frac{f_2\;(1+N_{\bf{p}})}{(\xi_2-\xi_1-\omega_{\bf{p}})^2}
    \;\right]\nonumber\\[3ex]
&+\cdots\nonumber
\end{align}
We perform the analytic continuation $ik_{\lambda} \rightarrow E+i\eta$, take the imaginary part of $1/(E- \xi_{2} \pm \omega_{\bf{p}} \pm \omega_{\bf{q}}+ i\eta)$, and get
\begin{align}
\textrm{Im}
\left\{
\frac{1}{\beta^2} \sum_{i\omega_{\kappa}}\sum_{i\omega_{\sigma}}
        f(i\omega_{\kappa},i\omega_{\sigma})\right\}
        \;=\; &
    -i\pi\delta(E- \xi_{2}- \omega_{\bf{p}}+ \omega_{\bf{q}})
        \frac{1}{(\xi_2-\xi_1+\omega_{\bf{p}})^2}
    \left[\;
        (1+N_{\bf{p}}-f_2)N_{\bf{q}}+N_{\bf{p}}f_2
    \;\right]\nonumber\\
&
    -i\pi\delta(E- \xi_{2}- \omega_{\bf{p}}- \omega_{\bf{q}})
        \frac{1}{(\xi_2-\xi_1+\omega_{\bf{p}})^2}
    \left[\;
        (1+N_{\bf{p}}-f_2)(1+N_{\bf{q}})-N_{\bf{p}}f_2
    \;\right]\nonumber\\
&
    -i\pi\delta(E- \xi_{2}+ \omega_{\bf{p}}+ \omega_{\bf{q}})
        \frac{1}{(\xi_2-\xi_1-\omega_{\bf{p}})^2}
    \left[\;
        (N_{\bf{p}}+f_2)N_{\bf{q}}+f_2(1+N_{\bf{p}})
    \;\right]\nonumber\\
&
    -i\pi\delta(E- \xi_{2}+ \omega_{\bf{p}}- \omega_{\bf{q}})
        \frac{1}{(\xi_2-\xi_1-\omega_{\bf{p}})^2}
    \left[\;
        (N_{\bf{p}}+f_2)(1+N_{\bf{q}})-f_2(1+N_{\bf{p}})
    \;\right]\nonumber\\
&+\cdots\nonumber
\end{align}
\pagebreak
\subsection{Two-Loop Diagram IIb}
The second two-loop diagram, called here IIb, is shown in the figure below. Since this diagram also has $L=2$ and $F=0$, the Feynman rules give
\begin{figure}[!h]
\includegraphics{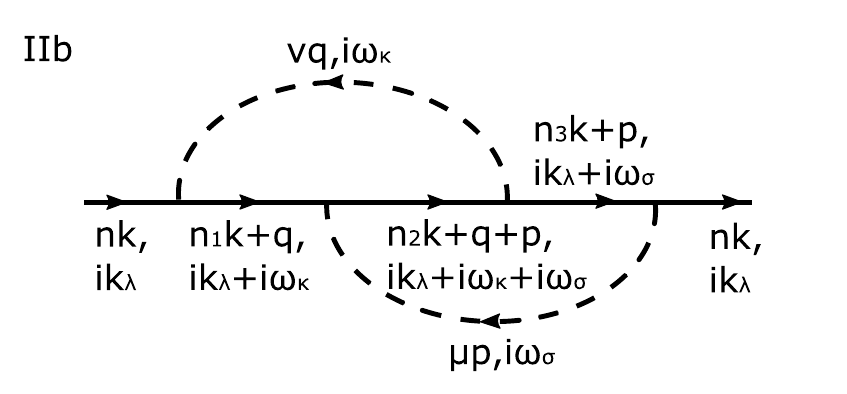}
%\caption{\label{fig:}}
\end{figure}
\begin{align}
%\overline
{\Sigma}^{\textrm{(IIb)}} = \frac{1}{\beta^{2} N_{\Omega}^{2}} &\;
        \sum_{n_{1}n_{2}n_{3}}
        \sum_{\nu\textbf{q}}
        \sum_{\mu\textbf{p}}
        g_{n_{1}n\nu}(\textbf{k},\textbf{q})
        g^{*}_{n_{2}n_{3}\nu}(\textbf{k}+\textbf{p},\textbf{q})
        g_{n_{2}n_{1}\mu}(\textbf{k}+\textbf{q},\textbf{p})
        g^{*}_{n_{3}n\mu}(\textbf{k},\textbf{p})
        \sum_{i\omega_{\kappa}}
        \sum_{i\omega_{\sigma}}
        f(i\omega_{\kappa},i\omega_{\sigma})
        \nonumber,
\end{align}
where $f(i\omega_{\kappa},i\omega_{\sigma})$ is defined as
\begin{align}
f(i\omega_{\kappa},i\omega_{\sigma}) \equiv
    &\; \frac{1}{ik_{\lambda}+i\omega_{\kappa}-\xi_{1}}\;
        \frac{1}{ik_{\lambda}+i\omega_{\kappa}
            +i\omega_{\sigma}-\xi_{2}} \;
        \frac{1}{ik_{\lambda}+i\omega_{\sigma}-\xi_{3\bf{p}}}\;
        \frac{2\omega_{\textbf{q}}}
        {-\omega_{\kappa}^{2}-\omega_{\textbf{q}}^{2}}\;
        \frac{2\omega_{\textbf{p}}}
        {-\omega_{\sigma}^{2}-\omega_{\textbf{p}}^{2}}
        \nonumber\\
    \equiv &\;
        A_{\kappa}
        \frac{1}{ik_{\lambda}+i\omega_{\kappa}
            +i\omega_{\sigma}-\xi_{2}}\;
        \frac{1}{ik_{\lambda}+i\omega_{\sigma}-\xi_{3\bf{p}}}\;
        \frac{2\omega_{\textbf{p}}}
        {-\omega_{\sigma}^{2}-\omega_{\textbf{p}}^{2}}
        \nonumber.
\end{align}
Using a notation we introduced above, $A_{\kappa}$ collects all terms independent of $\omega_{\sigma}$, and we use again abbreviations introduced in the previous section, such as $\xi_{3\bf{p}}\equiv \xi_{n_{3}\textbf{k}+\textbf{p}}$, etc. Summing over $i\omega_{\sigma}$ first, we get
\begin{align}
-\frac{1}{\beta} \sum_{i\omega_{\sigma}}
        f(i\omega_{\kappa},i\omega_{\sigma}) = &\;
    f_{2}A_{\kappa}
        \frac{1}{i\omega_{\kappa}-\xi_{2}+\xi_{3\bf{p}}}
        \left(
        \frac{1}{ik_{\lambda}+i\omega_{\kappa}
            -\xi_{2}-\omega_{\textbf{p}}}
        -\frac{1}{ik_{\lambda}+i\omega_{\kappa}
            -\xi_{2}+\omega_{\textbf{p}}}
        \right)
        \nonumber\\
    & + f_{3\bf{p}}A_{\kappa}
        \frac{1}{i\omega_{\kappa}-\xi_{2}+\xi_{3\bf{p}}}
        \left(
        \frac{1}{ik_{\lambda}
            -\xi_{3\bf{p}}+\omega_{\textbf{p}}}
        -\frac{1}{ik_{\lambda}
            -\xi_{3\bf{p}}-\omega_{\textbf{p}}}
        \right)
        \nonumber\\
    & + N_{\textbf{p}}A_{\kappa}
        \frac{1}{ik_{\lambda}+i\omega_{\kappa}
            -\xi_{2}+\omega_{\textbf{p}}}\;
        \frac{1}{ik_{\lambda}
            -\xi_{3\bf{p}}+\omega_{\textbf{p}}}
      + (N_{\textbf{p}}+1)A_{\kappa}
        \frac{1}{ik_{\lambda}+i\omega_{\kappa}
            -\xi_{2}-\omega_{\textbf{p}}}\;
        \frac{1}{ik_{\lambda}
            -\xi_{3\bf{p}}-\omega_{\textbf{p}}}
        \nonumber.
\end{align}
We then sum over $i\omega_{\kappa}$, and collect the relevant terms for two-phonon scattering processes, which are given below:
\begin{align}
\frac{1}{\beta^2}
        \sum_{i\omega_{\kappa}}\sum_{i\omega_{\sigma}}
        f(i\omega_{\kappa},i\omega_{\sigma}) \;=\;&\nonumber\\ 
\frac{1}{ik_{\lambda}- \xi_{2}- \omega_{\bf{p}}+ \omega_{\bf{q}}}
    &\left[\; 
        \frac{-f_2}{\xi_2-\xi_1+\omega_{\bf{p}}}\;
        \frac{1}{ik_{\lambda}-\xi_{3\bf{p}}-\omega_{\bf{p}}}\;
        \frac{f_2 N_{\bf{p}}}{1-f_2+N_{\bf{p}}}
       +\frac{f_2}{ik_{\lambda}-\xi_1+\omega_{\bf{q}}}\;
        \frac{N_{\bf{q}}}{\xi_{3\bf{p}}-\xi_2+\omega_{\bf{q}}}
        \right.\nonumber\\[3ex]
     & \left.
       +\frac{1+N_{\bf{p}}}{\xi_2-\xi_1+\omega_{\bf{p}}}\;
        \frac{1}{ik_{\lambda}-\xi_{3\bf{p}}-\omega_{\bf{p}}}\;
        \frac{f_2 N_{\bf{p}}}{1-f_2+N_{\bf{p}}}
       +\frac{1+N_{\bf{p}}}{ik_{\lambda}-\xi_1+\omega_{\bf{q}}}\;
        \frac{N_{\bf{q}}}{ik_{\lambda}-\xi_{3\bf{p}}-\omega_{\bf{p}}}
    \;\right]\nonumber\\[3ex]
+\frac{1}{ik_{\lambda}- \xi_{2}- \omega_{\bf{p}}- \omega_{\bf{q}}}
    &\left[\; 
        \frac{f_2}{\xi_2-\xi_1+\omega_{\bf{p}}}\;
        \frac{1}{ik_{\lambda}-\xi_{3\bf{p}}-\omega_{\bf{p}}}\;
        \frac{f_2 N_{\bf{p}}}{1-f_2+N_{\bf{p}}}
       +\frac{f_2}{ik_{\lambda}-\xi_1-\omega_{\bf{q}}}\;
        \frac{1+N_{\bf{q}}}{\xi_{3\bf{p}}-\xi_2-\omega_{\bf{q}}}
        \right.\nonumber\\[3ex]
     & \left.
       -\frac{1+N_{\bf{p}}}{\xi_2-\xi_1+\omega_{\bf{p}}}\;
        \frac{1}{ik_{\lambda}-\xi_{3\bf{p}}-\omega_{\bf{p}}}\;
        \frac{f_2 N_{\bf{p}}}{1-f_2+N_{\bf{p}}}
       +\frac{1+N_{\bf{p}}}{ik_{\lambda}-\xi_1-\omega_{\bf{q}}}\;
        \frac{1+N_{\bf{q}}}
        {ik_{\lambda}-\xi_{3\bf{p}}-\omega_{\bf{p}}}
    \;\right]\nonumber\\[3ex]
+\frac{1}{ik_{\lambda}- \xi_{2}+ \omega_{\bf{p}}+ \omega_{\bf{q}}}
    &\left[\; 
        \frac{f_2}{\xi_2-\xi_1-\omega_{\bf{p}}}\;
        \frac{1}{ik_{\lambda}-\xi_{3\bf{p}}+\omega_{\bf{p}}}\;
        \frac{f_2(1+N_{\bf{p}})}{f_2+N_{\bf{p}}}
       -\frac{f_2}{ik_{\lambda}-\xi_1+\omega_{\bf{q}}}\;
        \frac{N_{\bf{q}}}{\xi_{3\bf{p}}-\xi_2+\omega_{\bf{q}}}
        \right.\nonumber\\[3ex]
     & \left.
       +\frac{N_{\bf{p}}}{\xi_2-\xi_1-\omega_{\bf{p}}}\;
        \frac{1}{ik_{\lambda}-\xi_{3\bf{p}}+\omega_{\bf{p}}}\;
        \frac{f_2(1+N_{\bf{p}})}{f_2+N_{\bf{p}}}
       +\frac{N_{\bf{p}}}{ik_{\lambda}-\xi_1+\omega_{\bf{q}}}\;
        \frac{N_{\bf{q}}}{ik_{\lambda}-\xi_{3\bf{p}}+\omega_{\bf{p}}}
    \;\right]\nonumber\\[3ex]
+\frac{1}{ik_{\lambda}- \xi_{2}+ \omega_{\bf{p}}- \omega_{\bf{q}}}
    &\left[\; 
        \frac{-f_2}{\xi_2-\xi_1-\omega_{\bf{p}}}\;
        \frac{1}{ik_{\lambda}-\xi_{3\bf{p}}+\omega_{\bf{p}}}\;
        \frac{f_2(1+N_{\bf{p}})}{f_2+N_{\bf{p}}}
       -\frac{f_2}{ik_{\lambda}-\xi_1-\omega_{\bf{q}}}\;
        \frac{1+N_{\bf{q}}}{\xi_{3\bf{p}}-\xi_2-\omega_{\bf{q}}}
        \right.\nonumber\\[3ex]
     & \left.
       +\frac{-N_{\bf{p}}}{\xi_2-\xi_1-\omega_{\bf{p}}}\;
        \frac{1}{ik_{\lambda}-\xi_{3\bf{p}}+\omega_{\bf{p}}}\;
        \frac{f_2(1+N_{\bf{p}})}{f_2+N_{\bf{p}}}
       +\frac{N_{\bf{p}}}{ik_{\lambda}-\xi_1-\omega_{\bf{q}}}\;
        \frac{1+N_{\bf{q}}}
        {ik_{\lambda}-\xi_{3\bf{p}}+\omega_{\bf{p}}}
    \;\right]\nonumber\\[3ex]
+\cdots\nonumber\qquad\qquad\qquad\;\;\;&
\end{align}
After performing the analytic continuation and taking the imaginary part, we get
\begin{align}
\textrm{Im}& \;
\left\{
\frac{1}{\beta^2} \sum_{i\omega_{\kappa}} \sum_{i\omega_{\sigma}}
        f(i\omega_{\kappa},i\omega_{\sigma})\right\}
        \;=\; \nonumber\\
&
    -i\pi\delta(E- \xi_{2}- \omega_{\bf{p}}+ \omega_{\bf{q}})
        \frac{1}{\xi_2-\xi_1+\omega_{\bf{p}}}\;
        \frac{1}{\xi_2-\xi_{3\bf{p}}-\omega_{\bf{q}}}
    \left[\;
        N_{\bf{q}}+N_{\bf{q}}N_{\bf{p}}+N_{\bf{p}}f_2-N_{\bf{q}}f_2
    \;\right]\nonumber\\
&
    -i\pi\delta(E- \xi_{2}- \omega_{\bf{p}}- \omega_{\bf{q}})
        \frac{1}{\xi_2-\xi_1+\omega_{\bf{p}}}\;
        \frac{1}{\xi_2-\xi_{3\bf{p}}+\omega_{\bf{q}}}
    \left[\;
        (1+N_{\bf{p}})(1+N_{\bf{q}})-f_2(1+N_{\bf{p}}+N_{\bf{q}})
    \;\right]\nonumber\\
&
    -i\pi\delta(E- \xi_{2}+ \omega_{\bf{p}}+ \omega_{\bf{q}})
        \frac{1}{\xi_2-\xi_1-\omega_{\bf{p}}}\;
        \frac{1}{\xi_2-\xi_{3\bf{p}}-\omega_{\bf{q}}}
    \left[\;
        N_{\bf{p}}N_{\bf{q}}+N_{\bf{q}}f_2+N_{\bf{p}}f_2+f_2
    \;\right]\nonumber\\
&
    -i\pi\delta(E- \xi_{2}+ \omega_{\bf{p}}- \omega_{\bf{q}})
        \frac{1}{\xi_2-\xi_1-\omega_{\bf{p}}}\;
        \frac{1}{\xi_2-\xi_{3\bf{p}}+\omega_{\bf{q}}}
    \left[\;
        N_{\bf{p}}+N_{\bf{q}}N_{\bf{p}}+N_{\bf{q}}f_2-N_{\bf{p}}f_2
    \;\right]\nonumber\\
&+\cdots\nonumber
\end{align}
\subsection{Two-Phonon Scattering Rates}
Collecting the contributions from diagrams IIa and IIb, using $\Gamma=-(2/\hbar)\textrm{Im}\Sigma$, and setting $E$ to the band energy $\xi_{n\textbf{k}}$, the scattering rate of the two-phonon processes becomes
\begin{align}
\Gamma^{\textrm{(2ph)}}_{n\textbf{k}} =
       \frac{2\pi}{\hbar} \frac{1}{N_{\Omega}^{2}}
       \sum_{n_1n_2n_3}
       \sum_{\textbf{q}\textbf{p}}\sum_{\nu\mu}
     & \left[
        \gamma^{\textrm{(i)}}
           \delta(\xi_{n\textbf{k}}
           -\xi_2-\omega_{\textbf{p}}+\omega_{\textbf{q}})
       +\gamma^{\textrm{(ii)}}
           \delta(\xi_{n\textbf{k}}
           -\xi_2-\omega_{\textbf{p}}-\omega_{\textbf{q}})
           \right.\nonumber\\
     & \left.+\gamma^{\textrm{(iii)}}
           \delta(\xi_{n\textbf{k}}
           -\xi_2+\omega_{\textbf{p}}+\omega_{\textbf{q}})
       +\gamma^{\textrm{(iv)}}
           \delta(\xi_{n\textbf{k}}
           -\xi_2+\omega_{\textbf{p}}-\omega_{\textbf{q}})
           \right],\nonumber
\end{align}
where we introduce the process amplitudes
\begin{align}
\gamma^{\textrm{(i)}} = &\;
      \left(
        N_{\bf{q}}+N_{\bf{q}}N_{\bf{p}}+N_{\bf{p}}f_2-N_{\bf{q}}f_2
      \right)
      \times\nonumber\\[3ex]
      &\;\;
        \frac{g_{n_1n\nu}{(\bf{k},\bf{q})}
              g_{n_2n_1\mu}(\bf{k}+\bf{q},\bf{p})}
        {\xi_2-\xi_1+\omega_{\bf{p}}}
      \left(
        \frac{g_{n_3n\nu}^{*}{(\bf{k},\bf{q})}
              g_{n_2n_3\mu}^{*}(\bf{k}+\bf{q},\bf{p})}
        {\xi_2-\xi_3+\omega_{\bf{p}}}
        +
        \frac{g_{n_2n_3\nu}^{*}{(\bf{k}+\bf{p},\bf{q})}
              g_{n_3n\mu}^{*}(\bf{k},\bf{p})}
        {\xi_2-\xi_{3\bf{p}}-\omega_{\bf{q}}}
      \right)
      \nonumber\\[3ex]
\gamma^{\textrm{(ii)}} = &\;
      \left[
        (1+N_{\bf{q}})(1+N_{\bf{p}}-f_2)-N_{\bf{p}}f_2
      \right]
      \times\nonumber\\[3ex]
      &\;\;
        \frac{g_{n_1n\nu}{(\bf{k},\bf{q})}
              g_{n_2n_1\mu}(\bf{k}+\bf{q},\bf{p})}
        {\xi_2-\xi_1+\omega_{\bf{p}}}
      \left(
        \frac{g_{n_3n\nu}^{*}{(\bf{k},\bf{q})}
              g_{n_2n_3\mu}^{*}(\bf{k}+\bf{q},\bf{p})}
        {\xi_2-\xi_3+\omega_{\bf{p}}}
        +
        \frac{g_{n_2n_3\nu}^{*}{(\bf{k}+\bf{p},\bf{q})}
              g_{n_3n\mu}^{*}(\bf{k},\bf{p})}
        {\xi_2-\xi_{3\bf{p}}+\omega_{\bf{q}}}
      \right)
      \nonumber\\[3ex]
\gamma^{\textrm{(iii)}} = &\;
      \left[
        N_{\bf{q}}(N_{\bf{p}}+f_2)+(1+N_{\bf{p}})f_2
      \right]
      \times\nonumber\\[3ex]
      &\;\;
        \frac{g_{n_1n\nu}{(\bf{k},\bf{q})}
              g_{n_2n_1\mu}(\bf{k}+\bf{q},\bf{p})}
        {\xi_2-\xi_1-\omega_{\bf{p}}}
      \left(
        \frac{g_{n_3n\nu}^{*}{(\bf{k},\bf{q})}
              g_{n_2n_3\mu}^{*}(\bf{k}+\bf{q},\bf{p})}
        {\xi_2-\xi_3-\omega_{\bf{p}}}
        +
        \frac{g_{n_2n_3\nu}^{*}{(\bf{k}+\bf{p},\bf{q})}
              g_{n_3n\mu}^{*}(\bf{k},\bf{p})}
        {\xi_2-\xi_{3\bf{p}}-\omega_{\bf{q}}}
      \right)
      \nonumber\\[3ex]
\gamma^{\textrm{(iv)}} = &\;
      \left(
        N_{\bf{p}}+N_{\bf{q}}N_{\bf{p}}+N_{\bf{q}}f_2-N_{\bf{p}}f_2
      \right)
      \times\nonumber\\[3ex]
      &\;\;
        \frac{g_{n_1n\nu}{(\bf{k},\bf{q})}
              g_{n_2n_1\mu}(\bf{k}+\bf{q},\bf{p})}
        {\xi_2-\xi_1-\omega_{\bf{p}}}
      \left(
        \frac{g_{n_3n\nu}^{*}{(\bf{k},\bf{q})}
              g_{n_2n_3\mu}^{*}(\bf{k}+\bf{q},\bf{p})}
        {\xi_2-\xi_3-\omega_{\bf{p}}}
        +
        \frac{g_{n_2n_3\nu}^{*}{(\bf{k}+\bf{p},\bf{q})}
              g_{n_3n\mu}^{*}(\bf{k},\bf{p})}
        {\xi_2-\xi_{3\bf{p}}+\omega_{\bf{q}}}
      \right).
      \nonumber
\end{align}
Now we restore the infinitesimal $i\eta$ for the intermediate propagators. A useful sanity check is that our finite temperature results should reduce to the zero temperature results in the $T\rightarrow 0$ limit, 
from which we check that the pole structure of the finite and zero temperature expressions are consistent with each other. At zero temperature, we can directly compute the scattering amplitude $\mathcal{M}$. The zero temperature Feynman rules \{see Eqs. (2.124) to (2.127) in Ref. \cite{Mahan2000}\} give
\begin{align}
\mathcal{M} \sim \frac{gg}{E-\xi+i\eta}.\nonumber
\end{align}
The scattering rate is proportional to the absolute square of the scattering amplitude:
\begin{align}
\Gamma \sim \left|\mathcal{M}\right|^2 \sim 
    \left| \frac{gg}{E-\xi+i\eta} \right|^2 \nonumber.
\end{align}
Therefore, we will insert the infinitesimals in a way that allows us to express the scattering rates in an absolute square form. To achieve this, first note that quantities such as $\bf{q}$, $\bf{p}$, $\nu$ and $\mu$ are dummy variables that are summed over or integrated, so we can rename them at will. Let us denote $(\nu\bf{q} \leftrightarrow \mu\bf{p})$ the term with its dummy variables swapped in the way indicated by the arrows, $\nu\leftrightarrow\mu$, $\bf{q}\leftrightarrow \bf{p}$, etc. Let us consider the processes with one phonon absorption and one phonon emission first, that is, the sum of terms (i) and (iv):
\begin{align}
\sum_{n_1n_3}&\;\left[
        \gamma^{\textrm{(i)}}
           \delta(\xi_{n\textbf{k}}
           -\xi_2-\omega_{\mu\textbf{p}}+\omega_{\nu\textbf{q}})
       +\gamma^{\textrm{(iv)}}
           \delta(\xi_{n\textbf{k}}
           -\xi_2+\omega_{\mu\textbf{p}}-\omega_{\nu\textbf{q}})
           \right]\nonumber\\[3ex]
     & = \sum_{n_1n_3}\left[
        \gamma^{\textrm{(i)}}
           \delta(\xi_{n\textbf{k}}
           -\xi_2-\omega_{\mu\textbf{p}}+\omega_{\nu\textbf{q}})
       +\gamma^{\textrm{(iv)}}{(\nu\bf{q} \leftrightarrow \mu\bf{p})}
           \delta(\xi_{n\textbf{k}}
           -\xi_{2}+\omega_{\mu\textbf{p}}-\omega_{\nu\textbf{q}})
           {(\nu\bf{q} \leftrightarrow \mu\bf{p})}
           \right]\nonumber\\[3ex]
     & \equiv 
        \;\gamma^{\textrm{(1e1a)}}\;
           \delta(\xi_{n\textbf{k}}
           -\xi_2-\omega_{\mu\textbf{p}}+\omega_{\nu\textbf{q}})
           \nonumber,
\end{align}
where
\begin{align}
\gamma^{\textrm{(1e1a)}} &\;= \sum_{n_1n_3}
        \left[ 
          \gamma^{\textrm{(i)}}
        + \gamma^{\textrm{(iv)}}(\nu\bf{q} \leftrightarrow \mu\bf{p})
        \right]\nonumber\\[3ex]
&=\left(
        N_{\bf{q}}+N_{\bf{q}}N_{\bf{p}}+N_{\bf{p}}f_2-N_{\bf{q}}f_2
      \right)
      \;
      \sum_{n_1}\left(
        \frac{g_{n_1n\nu}{(\bf{k},\bf{q})}
              g_{n_2n_1\mu}(\bf{k}+\bf{q},\bf{p})}
        {\xi_2-\xi_1+\omega_{\bf{p}}}
       +\frac{g_{n_1n\mu}{(\bf{k},\bf{p})}
              g_{n_2n_1\nu}(\bf{k}+\bf{p},\bf{q})}
        {\xi_2-\xi_{1\bf{p}}-\omega_{\bf{q}}}
      \right)
      \nonumber\\[3ex]
    &\;\;\;\;\;\times
      \sum_{n_3}\left(
        \frac{g_{n_3n\nu}^{*}{(\bf{k},\bf{q})}
              g_{n_2n_3\mu}^{*}(\bf{k}+\bf{q},\bf{p})}
        {\xi_2-\xi_3+\omega_{\bf{p}}}
        +
        \frac{g_{n_2n_3\nu}^{*}{(\bf{k}+\bf{p},\bf{q})}
              g_{n_3n\mu}^{*}(\bf{k},\bf{p})}
        {\xi_2-\xi_{3\bf{p}}-\omega_{\bf{q}}}
      \right)
      \nonumber\\
&=\left(
        N_{\bf{q}}+N_{\bf{q}}N_{\bf{p}}+N_{\bf{p}}f_2-N_{\bf{q}}f_2
      \right)
      \;\left|
      \sum_{n_1}\left(
        \frac{g_{n_1n\nu}{(\bf{k},\bf{q})}
              g_{n_2n_1\mu}(\bf{k}+\bf{q},\bf{p})}
        {\xi_2-\xi_1+\omega_{\bf{p}}+i\eta}
       +\frac{g_{n_1n\mu}{(\bf{k},\bf{p})}
              g_{n_2n_1\nu}(\bf{k}+\bf{p},\bf{q})}
        {\xi_2-\xi_{1\bf{p}}-\omega_{\bf{q}}+i\eta}
      \right)\right|^2
      \nonumber.
\end{align}
Since the expression is already in square form, we have inserted the $i\eta$ terms in a way that makes the expression become an absolute square. Using a similar approach for the process in which the electron emits two phonons,
\begin{align}
\gamma^{\textrm{(2e)}} &\;= \sum_{n_1n_3}\gamma^{\textrm{(ii)}}
          \nonumber\\[3ex]
&=\sum_{n_1n_3}\;\frac{1}{2}
        \left[ 
          \gamma^{\textrm{(ii)}}
        + \gamma^{\textrm{(ii)}}
        \right]\nonumber\\[3ex]
&=\sum_{n_1n_3}\;\frac{1}{2}
        \left[ 
          \gamma^{\textrm{(ii)}}
        + \gamma^{\textrm{(ii)}}
          {(\nu\bf{q} \leftrightarrow \mu\bf{p})}
        \right]\nonumber\\[3ex]
&=\frac{1}{2}\left[
        (1+N_{\bf{q}})(1+N_{\bf{p}}-f_2)-N_{\bf{p}}f_2
      \right]
      \;\left|
      \sum_{n_1}\left(
        \frac{g_{n_1n\nu}{(\bf{k},\bf{q})}
              g_{n_2n_1\mu}(\bf{k}+\bf{q},\bf{p})}
        {\xi_2-\xi_1+\omega_{\bf{p}}+i\eta}
       +\frac{g_{n_1n\mu}{(\bf{k},\bf{p})}
              g_{n_2n_1\nu}(\bf{k}+\bf{p},\bf{q})}
        {\xi_2-\xi_{1\bf{p}}+\omega_{\bf{q}}+i\eta}
      \right)\right|^2
      \nonumber,
\end{align}
and for the process in which the electron absorbs two phonons,
\begin{align}
\gamma^{\textrm{(2a)}} &\;= \sum_{n_1n_3}\gamma^{\textrm{(iii)}}
          \nonumber\\[3ex]
&=\sum_{n_1n_3}\;\frac{1}{2}
        \left[ 
          \gamma^{\textrm{(iii)}}
        + \gamma^{\textrm{(iii)}}
        \right]\nonumber\\[3ex]
&=\sum_{n_1n_3}\;\frac{1}{2}
        \left[ 
          \gamma^{\textrm{(iii)}}
        + \gamma^{\textrm{(iii)}}
          {(\nu\bf{q} \leftrightarrow \mu\bf{p})}
        \right]\nonumber\\[3ex]
&=\frac{1}{2}\left[
        N_{\bf{q}}(N_{\bf{p}}+f_2)+(1+N_{\bf{p}})f_2
      \right]
      \;\left|
      \sum_{n_1}\left(
        \frac{g_{n_1n\nu}{(\bf{k},\bf{q})}
              g_{n_2n_1\mu}(\bf{k}+\bf{q},\bf{p})}
        {\xi_2-\xi_1-\omega_{\bf{p}}+i\eta}
       +\frac{g_{n_1n\mu}{(\bf{k},\bf{p})}
              g_{n_2n_1\nu}(\bf{k}+\bf{p},\bf{q})}
        {\xi_2-\xi_{1\bf{p}}-\omega_{\bf{q}}+i\eta}
      \right)\right|^2
      \nonumber.
\end{align}
We thus get:
\begin{align}
\Gamma^{\textrm{(2ph)}}_{n\textbf{k}} =
       \frac{2\pi}{\hbar} \frac{1}{N_{\Omega}^{2}}
       \sum_{n_2}
       \sum_{\textbf{q}\,\textbf{p}}\sum_{\nu\mu}
       \left[
        \gamma^{\textrm{(1e1a)}}
           \delta(\xi_{n\textbf{k}}
           -\xi_2-\omega_{\textbf{p}}+\omega_{\textbf{q}})
       +\gamma^{\textrm{(2e)}}
           \delta(\xi_{n\textbf{k}}
           -\xi_2-\omega_{\textbf{p}}-\omega_{\textbf{q}})
       +\gamma^{\textrm{(2a)}}
           \delta(\xi_{n\textbf{k}}
           -\xi_2+\omega_{\textbf{p}}+\omega_{\textbf{q}})
       \right].\nonumber
\end{align}
\subsection{Resonance}
The last expression is very close to the final result given in Eqs. (1)$-$(4) of the main text. The last problem we need to solve is that the sum giving $\Gamma^{\textrm{(2ph)}}_{n\bf{k}}$ in the expression above diverges when the intermediate electron state is on shell, in which case the denominator in the $\gamma$ terms given above vanishes, resulting in a divergent scattering rate. This phenomenon is called resonance. The problem is that the intermediate state will eventually transition into a different state, but using free propagators $1/(E-\xi+i\eta)$ for the on shell intermediate states implies an infinite intermediate state lifetime. The common practice in this situation, which also arises in other quantum field theories, is to consider the full electron propagator $1/(E-\xi+i\eta-\Sigma)$ as shown in the figure below, which introduces a finite lifetime for the intermediate electronic state. Diagramatically, this approach is equivalent to performing a resummation of diagrams to all orders, as is done in the well-known GW self-energy \{see Eq. (5.54) in Ref. \citep{Mahan2000}\}. For our 2ph scattering rate expression, we simply add the intermediate state self-energy in the denominators of all the $\gamma$ terms above, which removes the divergences.
\begin{figure}[h]
\includegraphics{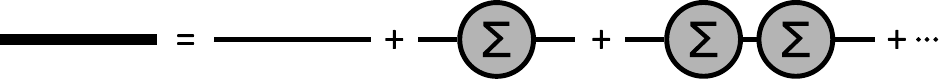}
%\caption{\label{fig:}}
\end{figure}
\subsection{Summary}
We rewrite the expression in more compact form. Defining the momentum of the final electronic state as $\textbf{k}^{\prime}\equiv\textbf{k}+\textbf{q}+\textbf{p}$, and using the following constants
\begin{align}
\alpha_{\bf{p}}^{\textrm{(1e1a)}}= \;\;\: 1,\;\;
\alpha_{\bf{p}}^{\textrm{(2e)}}=  1,\;\;
\alpha_{\bf{p}}^{\textrm{(2a)}}= -1,\;\;
\alpha_{\bf{q}}^{\textrm{(1e1a)}}= -1,\;\;
\alpha_{\bf{q}}^{\textrm{(2e)}}=  1,\;\;
\alpha_{\bf{q}}^{\textrm{(2a)}}= -1, \nonumber
\end{align}
we can write
\begin{align}
\Gamma^{\textrm{(2ph)}}_{n\textbf{k}} =
       \frac{2\pi}{\hbar} \frac{1}{N_{\Omega}^{2}}\;
       \sum_{n_2}\;
       \sum_{\nu\textbf{q}}\;
       \sum_{\mu\textbf{p}}
       \left[\;
           \widetilde{\Gamma}^{\textrm{(1e1a)}}
          +\widetilde{\Gamma}^{\textrm{(2e)}}
          +\widetilde{\Gamma}^{\textrm{(2a)}}
       \right],
       \nonumber
\end{align}
where
\begin{align}
\widetilde{\Gamma}^{\textrm{(i)}} = 
           \gamma^{\textrm{(i)}}\;
           \delta (
               \xi_{n\textbf{k}}
               -\xi_{n_2\textbf{k}^{\prime}}
               -\alpha_{\bf{p}}^{\textrm{(i)}} 
                   \omega_{\mu\textbf{p}}
               -\alpha_{\bf{q}}^{\textrm{(i)}} 
                   \omega_{\nu\textbf{q}}
           ).
           \nonumber
\end{align}
The square amplitudes $\gamma^{\textrm{(i)}}$ for the different processes, $\textrm{i}=\textrm{1e1a}$, $\textrm{2e}$ and $\textrm{2a}$, are defined as
\begin{align}
\label{eq:lowercase_gamma_abc}
\gamma^{\textrm{(i)}} =  A^{\textrm{(i)}} \;
   \left|
   \sum_{n_1}
   \left(
       \frac{g_{n_1n\nu}{(\bf{k},\bf{q})}
             g_{n_2n_1\mu}(\bf{k}+\bf{q},\bf{p})}
       {\xi_{n_2\textbf{k}^{\prime}}
        -\xi_{n_{1}\textbf{k}+\textbf{q}}
        +\alpha_{\bf{p}}^{\textrm{(i)}}\omega_{\mu\bf{p}}
        +i\eta
        -\Sigma_{n_{1}\textbf{k}+\textbf{q}}
        }
  +
       \frac{g_{n_1n\mu}{(\bf{k},\bf{p})}
             g_{n_2n_1\nu}(\bf{k}+\bf{p},\bf{q})}
       {\xi_{n_2\textbf{k}^{\prime}}
        -\xi_{n_{1}\textbf{k}+\textbf{p}}
        +\alpha_{\bf{q}}^{\textrm{(i)}}\omega_{\nu\bf{q}}
        +i\eta
        -\Sigma_{n_{1}\textbf{k}+\textbf{p}}
        }
   \right)\right|^2,
\end{align}
where we have taken into account the resonance by adding the intermediate state self-energy in the denominators. The factors of $A^{\textrm{(i)}}$ contain the thermal occupation numbers of electrons and phonons, and are defined as
\begin{align}
\label{eq:thermal_factor_A}
A^{\textrm{(1e1a)}} = &\; 
        N_{\nu\bf{q}}+N_{\nu\bf{q}}N_{\mu\bf{p}}
        +N_{\mu\bf{p}}f_{n_2\textbf{k}^{\prime}}
        -N_{\nu\bf{q}}f_{n_2\textbf{k}^{\prime}}
    \nonumber,
    \\[2ex]
A^{\textrm{(2e)}} = &\; 
    \frac{1}{2}\left[
        (1+N_{\nu\bf{q}})
        (1+N_{\mu\bf{p}}-f_{n_2\textbf{k}^{\prime}})
        -N_{\mu\bf{p}}f_{n_2\textbf{k}^{\prime}} 
    \right],
    \\[2ex]
A^{\textrm{(2a)}} = &\; 
    \frac{1}{2}\left[
        N_{\nu\bf{q}}
        (N_{\mu\bf{p}}+f_{n_2\textbf{k}^{\prime}})
        +(1+N_{\mu\bf{p}})f_{n_2\textbf{k}^{\prime}}
    \right]
    \nonumber.
\end{align}
These are our final expressions for the two-phonon scattering rates, which are given in Eqs. (1)$-$(4) of the main text.
\section{Temperature Dependence of the Two-Phonon Scattering Rates}
\indent 
Figure \ref{fig:T_dependence} below shows the temperature dependence of the ratios of the 2ph scattering rate and the leading order scattering rate, $\Gamma^{\mathrm{(2ph)}}/ \Gamma^{\mathrm{(1ph)}}$. Results are given for three electronic states, one in each of the regions I, II and III defined in the main text. 
The ratios are give for both the total 2ph scattering rate and (for completeness) for the individual 2ph processes, 1e1a, 2e and 2a. 
In the 200$-$500 K temperature range considered in this work, the ratio of the total 2ph scattering rate to the 1ph scattering rate is nearly temperature independent in all three energy regions.\\
%
% Figure 8
%
\begin{figure*}[!h]
\includegraphics[scale=1.0]{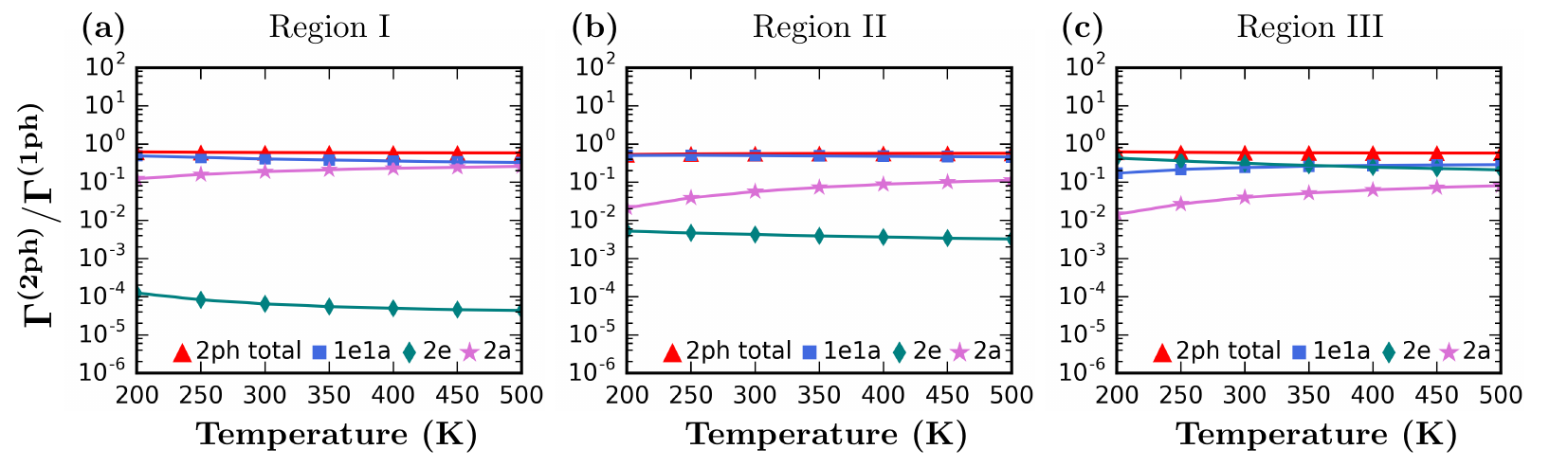}
\caption{Temperature dependence of the ratios of the 2ph scattering processes to the leading order $e$-ph scattering rate. From left to right, the panels are for electronic states with energies of 20, 45 and 90 meV above the conduction band minimum, and thus respectively in region I, II and III defined in the main text.} 
\label{fig:T_dependence}
\end{figure*}
\section{Boltzmann Transport Equation with Two-Phonon Contributions}
Let us briefly summarize the formulation of the linearized Boltzmann transport equation (BTE) incorporating the 2ph scattering processes. 
A more extensive explanation of its formulation and derivation will be presented in a separate work. 
Defining the total $e$-ph scattering rate as
\begin{align}
\Gamma_{n\bf{k}} 
=
\Gamma_{n\bf{k}}^{\textrm{(1ph)}}+\Gamma_{n\bf{k}}^{\textrm{(2ph)}}
\equiv
\frac{1}{N_{\Omega}}\sum_{m}\sum_{\nu\bf{q}} 
\widetilde{\Gamma}_{n\bf{k},\;\nu\bf{q}}^{\textrm{(1ph)}}
+
\frac{1}{N_{\Omega}^2}\sum_{n_2}\sum_{\nu\bf{q}}\sum_{\mu\bf{p}}
\widetilde{\Gamma}_{n\bf{k},\;\nu\bf{q},\;\mu\bf{p}}^{\textrm{(2ph)}}
\nonumber\,\,,
\end{align}
the linearized BTE can be expressed as
\begin{align}
\label{eq:BTE}
\textbf{F}_{n\bf{k}} = \textbf{F}^0_{n\bf{k}}
+\tau_{n\bf{k}} 
  \left[
  \frac{1}{N_{\Omega}}\sum_{m}\sum_{\nu\bf{q}}  
    \textbf{F}_{m\bf{k}+\bf{q}}
    \widetilde{\Gamma}_{n\bf{k},\;\nu\bf{q}}^{\textrm{(1ph)}}
  +
    \frac{1}{N_{\Omega}^2}\sum_{n_2}\sum_{\nu\bf{q}}\sum_{\mu\bf{p}}
    \textbf{F}_{n_2\bf{k}^{\bf{\prime}}}
    \widetilde{\Gamma}_{n\bf{k},\;\nu\bf{q},\;\mu\bf{p}}^{\textrm{(2ph)}}
  \right],
\end{align}
where the relaxation time $\tau_{n\bf{k}}$ is the inverse of the total scattering rate, namely $\tau_{n\bf{k}} = 1/\Gamma_{n\bf{k}}$, and the first and second terms in brackets are due to the lowest-order (1ph) and 2ph scattering processes, respectively. The function $\textbf{F}_{n\bf{k}}$ is the unknown in the equation, and $\textbf{F}^0_{n\bf{k}} = \tau_{n\bf{k}} \textbf{v}_{n\bf{k}}$, with $\textbf{v}_{n\bf{k}}$ the band velocity. After solving the equation, the electrical mobility in direction $i$ can be obtained using
\begin{align}
\mu^i = 
\frac{2 e \beta}{n_c V_{\textrm{uc}}}
\sum_{n\bf{k}} f_{n\bf{k}} (1-f_{n\bf{k}})
v^i_{n\bf{k}} F^i_{n\bf{k}}
\nonumber,
\end{align}
where $V_{\textrm{uc}}$ is the unit cell volume and $n_c$ the charge carrier concentration.\\
\indent 
A common approach to computing $\textbf{F}_{n\bf{k}}$ is the relaxation time approximation (RTA), which neglects the second term on the right hand side of Eq. (\ref{eq:BTE}) and approximates $\textbf{F}_{n\bf{k}}$ to $\textbf{F}^0_{n\bf{k}}$. A more accurate solution can be obtained through an iterative approach (ITA)~\cite{Li2015}, in which, starting from the RTA solution $\textbf{F}_{n\bf{k}}=\textbf{F}^0_{n\bf{k}}$, one iteratively substitutes $\textbf{F}_{n\bf{k}}$ in the right hand side of Eq. (\ref{eq:BTE}) to obtain an updated $\textbf{F}_{n\bf{k}}$, until a converged $\textbf{F}_{n\bf{k}}$ is reached.\\
\indent 
Our work presents calculations, both within the RTA and ITA, in which the 2ph processes are either included or neglected; when only 1ph processes are included, 
$\tau_{n\bf{k}}$ is set to the inverse of $\Gamma_{n\bf{k}}^{\mathrm{(1ph)}} $, and the second term in brackets in Eq.~(\ref{eq:BTE}) is neglected. 
The ITA with 2ph contributions included is the most accurate level of theory, and the one that agrees best with experiment, while the ITA with only 1ph processes overestimates the experimental result.

\bibliography{supplemental_materials}